\documentclass[twocolumn]{aastex631}
\usepackage{hyperref}
\usepackage{soul}
\usepackage{todonotes}
\usepackage{multirow}
\usepackage{color, colortbl}
\usepackage{tabularx}
\definecolor{Gray}{gray}{0.95}
\usepackage{graphicx}
\usepackage{booktabs}
\usepackage{rotating}
\usepackage{amsmath}
\usepackage{amssymb}
\usepackage{comment}
\usepackage{subfigure}

\newcommand{\icarogw}{\texttt{icarogw}}
\newcommand{\bilby}{\texttt{Bilby}}

\shorttitle{GW Cosmology Blinded MDC-I}
\begin{document}

\title{Blinded Mock Data Challenge for Gravitational-Wave Cosmology-I: Assessing the Robustness of Methods Using Binary Black Holes Mass Spectrum}
\author[0000-0002-8685-5477]{Aman Agarwal}\altaffiliation{Editorial Team and Analysis Group}\affiliation{Perimeter Institute for Theoretical Physics, Waterloo, Ontario, N2L 2Y5, Canada }\affiliation{Department of Physics, University of Guelph, Guelph, Ontario, N1G 2W1, Canada}\affiliation{Institute of Physics, University of Greifswald, 17489 Greifswald, Germany}
\author[0000-0003-2766-247X]{Ulyana Dupletsa}\altaffiliation{Editorial Team and Analysis Group}\affiliation{Gran Sasso Science Institute (GSSI), I-67100 L'Aquila, Italy}
\affiliation{INFN, Laboratori Nazionali del Gran Sasso, I-67100 Assergi, Italy}\affiliation{Institute of High Energy Physics - Austrian Academy of Sciences, 1010 Vienna, Austria}
\author[0000-0001-7661-2810]{Konstantin Leyde}\altaffiliation{Editorial Team and Analysis Group}\affiliation{ICG Portsmouth,University of Portsmouth: Portsmouth, Hampshire, United Kingdom}
\author[0000-0002-3373-5236]{Suvodip Mukherjee}\altaffiliation{Coordinator, Editorial Team, and Simulation Group}\affiliation{Department of Astronomy and Astrophysics, Tata Institute of Fundamental Research, Mumbai 400005, India}
\author[0000-0002-7629-4805]{Beno\^it Revenu}\altaffiliation{Editorial Team and Simulation Group}\affiliation{Subatech, CNRS - Institut Mines-Telecom Atlantique - Universit\'e de Nantes, France}\affiliation{Université Paris-Saclay, CNRS/IN2P3, IJCLab, 91405 Orsay, France}
\author{Juan Esteban Rivera}\altaffiliation{Editorial Team and Simulation Group}\affiliation{Instituto de Fisica, Universidad de Antioquia, A.A.1226, Medellin, Colombia}
\author[0000-0002-0314-8698]{Antonio Enea Romano}\altaffiliation{Editorial Team and Simulation Group}\affiliation{Instituto de Fisica, Universidad de Antioquia, A.A.1226, Medellin, Colombia}
\author[0009-0005-9881-1788]{Mohit Raj Sah}\altaffiliation{Editorial Team and Simulation Group}\affiliation{Department of Astronomy and Astrophysics, Tata Institute of Fundamental Research, Mumbai 400005, India}
\author[0000-0002-6827-9509]{Sergio Vallejo-Pe\~na}\altaffiliation{Editorial Team and Simulation Group}\affiliation{Instituto de Fisica, Universidad de Antioquia, A.A.1226, Medellin, Colombia}
\author{Adrian Avendano}\altaffiliation{Simulation Group}\affiliation{Instituto de Fisica, Universidad de Antioquia, A.A.1226, Medellin, Colombia}
\author[0000-0002-4003-7233]{Freija Beirnaert}\altaffiliation{Simulation Group}\affiliation{Universiteit Gent, Proeftuinstraat 86, 9000 Ghent, Belgium}
\author[0000-0003-3258-5763]{Gergely Dalya}\altaffiliation{Simulation Group}\affiliation{Laboratoire des 2 Infinis - Toulouse (L2IT-IN2P3), Universit\'e de Toulouse, CNRS, UPS, F-31062 Toulouse Cedex 9, France}\affiliation{
MTA-ELTE Astrophysics Research Group, 1117 Budapest, Hungary}
\author[0009-0006-1111-3213]{Miguel Cifuentes Espitia}\altaffiliation{Analysis Group}\affiliation{Instituto de Fisica, Universidad de Antioquia, A.A.1226, Medellin, Colombia}
\author[0000-0002-0642-5507]{Christos Karathanasis}\altaffiliation{Simulation Group}\affiliation{Institut de Física d’Altes Energies (IFAE), Barcelona Institute of Science and Technology, Barcelona, Spain}
\author[0009-0009-8350-3247]{Santiago Moreno-Gonzalez}\altaffiliation{Analysis Group}\affiliation{Instituto de Fisica, Universidad de Antioquia, A.A.1226, Medellin, Colombia}
\author[0009-0009-2287-2102]{Lucas Quiceno}\altaffiliation{Analysis Group}\affiliation{Instituto de Fisica, Universidad de Antioquia, A.A.1226, Medellin, Colombia}
\author[0000-0002-8658-5753]{Federico Stachurski}\altaffiliation{Simulation Group}\affiliation{SUPA, University of Glasgow, Glasgow, G12 8QQ, United Kingdom}
\author[0000-0002-9370-8360]{Juan Garcia-Bellido}\affiliation{Instituto de Fisica Teorica IFT-UAM/CSIC, Universidad Autonoma de Madrid, 28049 Madrid, Spain}
\author[0000-0002-5556-9873]{Rachel Gray}\affiliation{SUPA, University of Glasgow, Glasgow, G12 8QQ, United Kingdom}
\author[0000-0001-8760-5421]{Nicola Tamanini}\affiliation{Laboratoire des 2 Infinis - Toulouse (L2IT-IN2P3), Universit\'e de Toulouse, CNRS, UPS, F-31062 Toulouse Cedex 9, France}
\author[0000-0001-7122-6240]{Cezary Turski}\affiliation{Universiteit Gent, Proeftuinstraat 86, 9000 Ghent, Belgium}
\correspondingauthor{Suvodip Mukherjee, \href{suvodip@tifr.res.in}{suvodip@tifr.res.in}}
\begin{abstract}
Gravitational Wave (GW) sources are standard sirens that provide an independent way to map the cosmic expansion history by combining with an independent redshift measurement either from an electromagnetic counterpart for a bright siren or using different statistical techniques for dark sirens. In this analysis, we perform the first Blinded Mock Data Challenge (\texttt{Blinded-MDC}) to test the robustness in inferring the value of Hubble constant $H_0$ for a dark siren technique which depends on astrophysical mass distribution of Binary Black Holes (BBHs). We have considered different analysis setups for the \texttt{Blinded-MDC} to test both statistical and systematic uncertainties and demonstrate the capabilities in inferring $H_0$ with detector sensitivity as per the fourth observation run of LIGO-Virgo-KAGRA. We find that when the astrophysical population of BBHs matches with the underlying assumption of the model, a cosmological pipeline can recover the injected parameters using the observed mass distribution. However, when the mock mass distribution of the astrophysical population depends slightly on redshift and one is ignorant about it in analyzing the data, it can cause a systematic discrepancy in the inferred value of $H_0$ by about $1.5\sigma$, above the statistical fluctuations due to noise and a limited number of events. In the future, elaborate studies will be required to mitigate systematic uncertainties due to unknown astrophysical complexities. This MDC framework sets the road map for inspecting the precision and accuracy of standard siren cosmology and provides the first insight into the robustness of the population-dependent cosmology inference in a blinded analysis setup.
\end{abstract}


\section{Introduction}
The detection of Gravitational Waves (GWs) by the LIGO-Virgo-KAGRA (LVK) Scientific Collaboration has opened a window to the cosmos, making it possible to explore physics from the smallest scale associated with neutron stars to cosmological scales of the Universe and answering several fundamental questions about the Universe \citep{PhysRevLett.116.061102}. One such fundamental question is the expansion history of the Universe as a function of cosmological redshift. Such a measurement will not only shed light on the mismatch in the value of the current expansion rate (called the Hubble constant $H_0$)\citep{Verde:2019ivm,Abdalla:2022yfr} but will also discover the constituents that contribute significantly to the energy budget of the Universe such as dark matter and dark energy. 

Measurement of the expansion history of the Universe using GW sources is feasible using sources having electromagnetic (EM) counterparts such as binary neutron stars (BNSs) and neutron star-black holes (NSBHs), and sources without EM counterparts such as stellar origin binary black holes (BBHs)  {without the presence of any baryonic matter around it. However, the population of BBHs formed in  Active Galactic Nucleus (AGN) disc can have EM counterparts \citep{ Gayathri:2020mra, Mukherjee:2020kki, Chen:2020gek,Bom:2023zgw}.
 }   For the former, measuring the redshift of the host galaxy of the GW sources is possible through spectroscopic follow-ups of the EM counterpart, as shown in the case of GW170817 \citep{TheLIGOScientific:2017qsa, LIGOScientific:2017adf}. For the latter, various techniques have been developed and applied to the second (GWTC-2) or third (GWTC-3) GW catalog of the LVK Scientific Collaboration \citep{LIGOScientific:2019zcs, Finke:2021aom, LIGOScientific:2021aug, Mukherjee:2022afz,Karathanasis:2022rtr}. These techniques include a mass spectrum of BBHs through spectral sirens technique  \citep{Farr:2019twy, Mastrogiovanni:2021wsd, Mukherjee:2021rtw, Leyde:2022orh, Ezquiaga:2022zkx, Mastrogiovanni:2023zbw, Leyde:2023iof, Pierra:2023deu, MaganaHernandez:2024uty, Farah:2024xub, Mali:2024wpq}, statistical host identification technique using galaxy catalogs \citep{Schutz, Fishbach:2018gjp,Soares-Santos:2019irc,PhysRevD.101.122001, Soares-Santos:2019irc, Palmese:2021mjm, Gray:2023wgj}, and the GW-galaxy cross-correlation technique for LVK \citep{Mukherjee:2018ebj,Mukherjee:2019wcg,Mukherjee:2020hyn, Bera:2020jhx,Ferri:2024amc} which will be useful in synergy with the spectroscopic galaxy surveys such as DESI \citep{Aghamousa:2016zmz}, Euclid \citep{2010arXiv1001.0061R}, and SPHEREx\citep{Dore:2018kgp} and upcoming photometric survey Vera Rubin Observatory \citep{LSSTScienceBook} as demonstrated in \citep{Diaz:2021pem,Afroz:2024joi}.  {Using  these techniques, the value of the  Hubble constant after combining both bright and dark standard sirens inferred using spectral sirens assuming no redshift evolution of mass distribution is $H_0= 68^{+12}_{-8}$ km s$^{-1}$Mpc$^{-1}$ \citep{LIGOScientific:2021aug}, using statistical host identification assuming a fixed BBH population model is $H_0= 68^{+8}_{-6}$ km s$^{-1}$Mpc$^{-1}$\citep{LIGOScientific:2021aug}, and using the cross-correlation technique (after marginalizing over BBH bias parameters controlling the population assumption) is $H_0= 75.4^{+11}_{-6}$ km s$^{-1}$Mpc$^{-1}$ \citep{Mukherjee:2022afz}.}   

However, one of the key requirements for an accurate and precise cosmological probe is to understand the impact of astrophysical uncertainty due to various assumptions and the reliability of a technique in providing accurate cosmological results despite these challenges. Such analyses are performed for other cosmological probes to the expansion history such as Cosmic Microwave Background (CMB) \citep{Planck:2015txa}, Baryon Acoustic Oscillation (BAO) \citep{DESI:2024ude}, and lensing time delay measurements \citep{Ding:2018hai} to understand the impact from possible known sources of systematics. In this work, we explore for the first time a \textit{Blinded Mock Data Challenge} (\texttt{Blinded-MDC}) on realistically simulated GW sources to understand the interplay between cosmological inference of $H_0$ and astrophysical population assumptions of BBHs. This \texttt{Blinded-MDC} analysis is focused on the method that uses the mass distribution of the BBHs to infer the expansion history of the Universe and the impact of the underlying assumption on the spectral-sirens technique. In the future, the \texttt{Blinded-MDC} technique will be applied to other astrophysical scenarios for the spectral-siren method, as well as for other techniques to understand the robustness of the standard siren methods in inferring the cosmological parameters, primarily the Hubble parameter $H(z)$. Apart from systematic errors due to astrophysical populations of BBHs, the impact of a few other sources of systematic errors in GW cosmology such as inclination angle \citep{Muller:2024wzl, Salvarese:2024jpq}, inaccurate waveform \citep{PhysRevD.110.043520}, peculiar velocity of host galaxy \citep{Mukherjee:2019qmm, Nimonkar:2023pyt}, and photometric redshifts uncertainty of host galaxy \citep{Turski:2023lxq}, were explored previously. 

The paper is organized as follows: we outline the setup of the \texttt{Blinded-MDC} in section \ref{sec-mdc}. The simulated mock data for different astrophysical cases considered in this analysis are discussed in section \ref{sec-pop}. The formalism we have used for the analysis is discussed in section \ref{sec-formalism}.  The results and the discussion for different cases from the \texttt{Blinded-MDC} are presented in section \ref{sec-results}. Finally, the conclusion and future scopes are discussed in section \ref{sec-conc}. 

\section{Setup for the Blinded Mock Data Challenge}\label{sec-mdc}
The \texttt{Blinded-MDC} is set up to find out the robustness of a GW cosmology analysis pipeline in inferring the cosmological parameters in two aspects, namely (i) in the presence of statistical uncertainties of the GW source parameters without any modeling error in the astrophysical population of BBHs and (ii) in the presence of modeling error(s). {For this \texttt{Blinded-MDC} we have focused on two scenarios, (i) the vanilla case: where the underlying population model is \textsc{Power Law + Gaussian Peak} (PLG) without any redshift evolution \citep{KAGRA:2021duu}, but the values of the parameters in PLG (discussed in the next section, \ref{sec:vanilla_model}) and the values of the cosmological parameters are blinded, and (ii) the Redshift Dependent scenario: where the underlying population model is PLG with redshift-dependent parameters (as discussed in the next section, \ref{sec:redshift-dependent}). {This differs from the usual assumption of the no redshift evolution of the mass distribution of BBHs \citep{LIGOScientific:2019zcs, Mastrogiovanni:2023zbw, Gray:2023wgj}.}} 

The injected values are chosen from a wide prior range of the astrophysical and cosmological parameters and the \texttt{Blinded-MDC} is made blind to ensure that the choice in the analysis settings is independent of the injected values, to test the reliability of the pipelines and to avoid any confirmation bias. As a result, two disjoint research groups are prepared, (i) the simulation team, and (ii) the analysis team, which are coordinated by the MDC coordinator. The exact setup of the \texttt{Blinded-MDC} is shown by a schematic diagram in Fig. \ref{fig:mdcsetup}. We describe below each of the cases:

\begin{figure*}[]
    \centering
    \includegraphics[scale = 0.9]{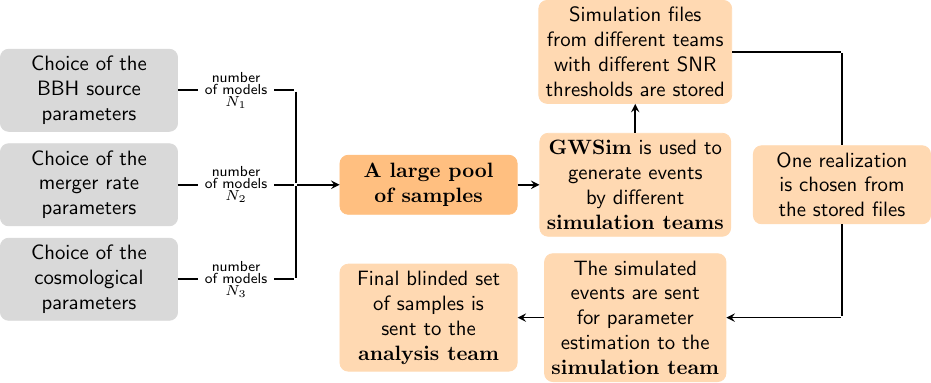}
    \caption{Schematic representation of the \texttt{Blinded-MDC} setup.}
    \label{fig:mdcsetup}
\end{figure*}

 {\texttt{Blinded-MDC} Simulation setup: } The code \texttt{GWSIM} \citep{Karathanasis:2022hrb} is used for the generation of the simulated GW mock data using the LVK O1+O2+O3+O4 noise power spectral densities (PSDs) for the vanilla case and with the O4 noise PSD for the Redshift-Dependent scenario \citep{LIGOScientific:2014pky, VIRGO:2014yos, PhysRevD.88.043007} with the duty cycles as mentioned in Tab.  \ref{tab:duty}. The values of the BBH source parameters, merger rate parameters, and cosmological parameters are drawn randomly from a fixed prior range before passing it to the simulation code \texttt{GWSIM}. The \texttt{GWSIM} code has gone through an internal review for the validation of the injected simulation set with the expected distribution for the model of the astrophysical population.  For the vanilla case, the simulation set that was passed to the analysis team is for the values for which the GWSIM code was validated. This is to ensure that there is no error in the pipeline of \texttt{Blinded-MDC} setup, which includes both injection and analysis parts. The values of the injected parameters and whether they are the same as the one for which the code is tested remained blinded until the analysis was completed. The matched filtering Signal to Noise Ratio (SNR) denoted by $\rho$ thresholds used for the Vanilla case is $\rho_{\rm th}= 10$. For the Redshift-Dependent scenario, the matched filtering SNR threshold is $\rho_{\rm th}= 12$.  As the Redshift-Dependent MDC scenario, explores the impact on $H_0$ due to the mis-modeling of the astrophysical population assumption, we have chosen a higher SNR threshold for this case to make sure that the contamination from noise fluctuation is not significant and hence impact from statistical fluctuation is limited on the error budget of the cosmological and astrophysical population. This helps in understanding more clearly the effect of systematic errors over the statistical uncertainties.  

The injection set is prepared for ten cases (denoted by I(j, $\rho_{\rm th}$), where the index j runs from $j=1$ to $j=10$) for each threshold with the values of the cosmological and astrophysical population parameters  {are drawn} randomly from a uniform distribution over the range by different members of the simulation team (at least by three members). Then a particular simulation from I($j'$, $\rho_{\rm th}$) is randomly chosen by the coordinator and a simulation team member for which estimation of the GW source parameters is performed using the package \bilby{} \citep{bilby_paper}. The particular realization (denoted by $j'$) chosen for the PE and the corresponding values of the cosmological and astrophysical parameters chosen are not known to anyone in the \texttt{Blinded-MDC} team until the end of the analysis by the analysis team.  {As for every different realization, the underlying population parameters and the cosmological parameters are different, a randomly chosen realization} makes the injection part blind to everyone, except the coordinator and one member of the simulation team. For the other simulations, \bilby{} parameter estimations were not performed to reduce the computation cost. However, these sets are available after the unblinding and can be used for any checking purposes. 
\begin{center}
    \begin{table}[h]
    \caption{List of detectors with the duty cycle used in the analysis. The sensitivity considered for the analysis corresponds to the relative observing run.}
        \label{tab:duty}
\hspace{-2cm}\begin{tabular}{| c | c || c | c || c |c|}
\hline
\multicolumn{6}{|c|}{For the Vanilla case with $T_{\rm obs}= 3$ years.}\\
\hline
\multicolumn{2}{|c|}{Detector} & \multicolumn{4}{|c|}{Observation Run}\\
\hline
\multicolumn{2}{|c|}{} & \multicolumn{1}{|c|}{O1} & \multicolumn{1}{|c|}{O2} & \multicolumn{1}{|c|}{O3} & \multicolumn{1}{|c|}{O4}\\
\hline
\multicolumn{2}{|c|}{H}&\multicolumn{1}{|c|}{0.6}&\multicolumn{1}{|c|}{0.6}&\multicolumn{1}{|c|}{0.75}&\multicolumn{1}{|c|}{0.75}\\
\hline
\multicolumn{2}{|c|}{L}&\multicolumn{1}{|c|}{0.5}&\multicolumn{1}{|c|}{0.6}&\multicolumn{1}{|c|}{0.75}&\multicolumn{1}{|c|}{0.75}\\
\hline
\multicolumn{2}{|c|}{V}&\multicolumn{1}{|c|}{--}&\multicolumn{1}{|c|}{--}&\multicolumn{1}{|c|}{0.75}&\multicolumn{1}{|c|}{0.75}\\
\hline
\hline
\multicolumn{6}{|c|}{For the Redshift-Dependent scenario with $T_{\rm obs}= 1$ year.}\\
\hline
\multicolumn{2}{|c|}{Detector} & \multicolumn{4}{|c|}{Observation Run}\\
\hline
\multicolumn{2}{|c|}{} & \multicolumn{1}{|c|}{O1} & \multicolumn{1}{|c|}{O2} & \multicolumn{1}{|c|}{O3} & \multicolumn{1}{|c|}{O4}\\
\hline
\multicolumn{2}{|c|}{H}&\multicolumn{1}{|c|}{--}&\multicolumn{1}{|c|}{--}&\multicolumn{1}{|c|}{--}&\multicolumn{1}{|c|}{0.75}\\
\hline
\multicolumn{2}{|c|}{L}&\multicolumn{1}{|c|}{--}&\multicolumn{1}{|c|}{--}&\multicolumn{1}{|c|}{--}&\multicolumn{1}{|c|}{0.75}\\
\hline
\multicolumn{2}{|c|}{V}&\multicolumn{1}{|c|}{--}&\multicolumn{1}{|c|}{--}&\multicolumn{1}{|c|}{--}&\multicolumn{1}{|c|}{0.75}\\
\hline
\end{tabular}
    \end{table}
\end{center}

 {\texttt{Blinded-MDC} analysis setup: } The code \texttt{icarogw} \citep{Mastrogiovanni:2023zbw} is used for the analysis purpose of exploring the interplay between the cosmological parameters and the population parameters. {The analysis team used a realization given by the simulation team of the posteriors on the GW source parameters ($\mathcal{M}$, q, $d_L$, Dec, RA, $i$, $\Psi$, $\phi$, and $t_c$) without any information about the cosmological and astrophysical population parameters used to generate them.} Using this set the analysis team performs the joint inference of the cosmological parameters and the population parameters for the underlying model considered as the fiducial astrophysical model PLG  without redshift evolution and Flat LCDM cosmological model. The analysis team explores the impact of different choices such as the number of samples in the selection function (or in the injection set), the impact of wider priors, and the impact on the inference with a change in the number of detected events. Finally, once the analysis members are satisfied with all the checks, the values of the cosmological and the population parameters used for the simulations get unblinded and are compared with the inferred values. 

The Vanilla case allows us to verify the analysis setup on whether the analysis code can infer the values correctly for a known scenario. For the second case i.e. {Redshift-Dependent} scenario, the \texttt{Blinded-MDC} inspects the impact on the inference of the parameters (both astrophysical and cosmological), when the underlying true astrophysical model differs from the fiducial assumption of the model. In the remainder of the paper, we describe in detail the results from the injection and the analysis parts.  

\section{Cases studied in the \texttt{Blinded-MDC}}\label{sec-pop}
As introduced in section \ref{sec-mdc}, the scope of the \texttt{Blinded-MDC} is twofold, in the first place, we want to quantify statistical uncertainties under the assumption that the settings used to generate and analyze data are the same. This is the Vanilla case. It allows us to assess the impact due to detector noise and the fact that our data represent a limited sample. Specifically, we want to study the presence of statistical fluctuations and outliers among data that might bias the overall results.  {Secondly, our goal is to evaluate systematic errors arising from astrophysical mis-modeling. To do this, we will only assume a redshift dependence in the data generation process, not in the analysis.} We address this in the Redshift-Dependent scenario. We provide a detailed explanation of the astrophysical models used to create the events dataset for both scenarios. 

In this work, the GW source parameter estimation for the simulated mock events is performed with \bilby{} \citep{bilby_paper} using the \texttt{IMPRhenomPv2} waveform approximant \citep{waveform1,waveform2,waveform3}, using the standard \bilby{} priors for BBHs with all the spin parameters fixed to zero (delta-function prior). An example event analysis is shown in the appendix in Fig.~\ref{fig:bilby_vanilla}. The analysis is done on the  {following} $9$ parameters: the chirp mass ${\mathcal{M}}$ (a combination of the two-component masses in detector frame\footnote{The chirp mass is defined as in terms of component masses ($m_1, m_2$): \[{\mathcal{M}} \equiv \frac{(m_1 m_2)^\frac{3}{5}}{(m_1 + m_2)^\frac{1}{5}}\]}), the mass ratio $q$ (defined as the ratio between the lightest and the heaviest mass component), the luminosity distance $d_L$ to the source, the sky position given by the right ascension RA and declination Dec, the inclination angle $\iota$ (defined as the angle between the line of sight and the perpendicular to the orbital plane of the binary), the polarization angle $\Psi$, the phase $\phi$ of the GW signal and the time of coalescence $t_c$. The choice of the non-spinning waveform will not have any noticeable impact on the cosmology results obtained in this paper for this detector noise.   

\subsection{Vanilla model}
\label{sec:vanilla_model}
\begin{figure}
  \subfigure[]{\label{fig:Pd1}
    \centering
    \includegraphics[width=\linewidth,trim={0.cm 0  0 0.cm},clip]{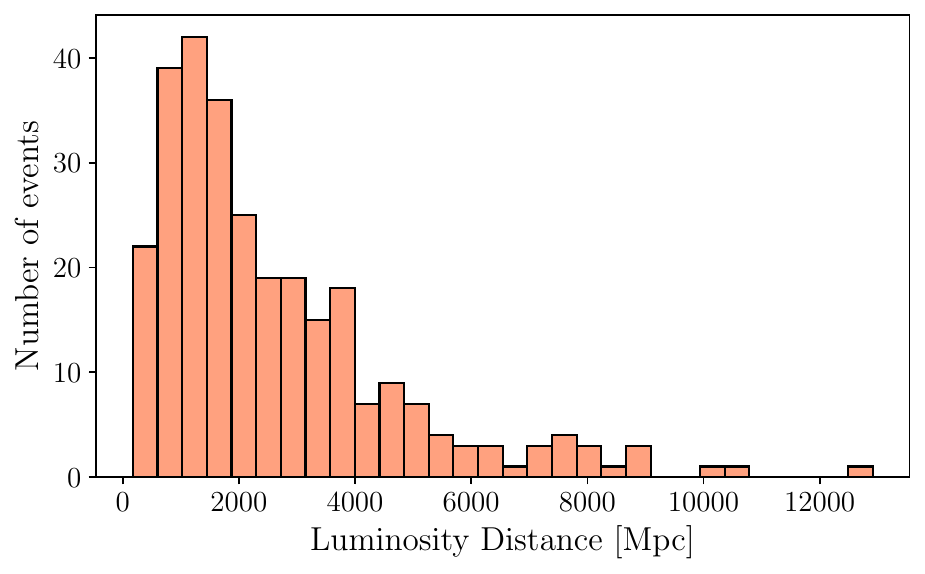}}
  \subfigure[]{\label{fig:Pm1}
    \centering
    \includegraphics[width=\linewidth,trim={0.cm 0cm  0cm 0.cm},clip]{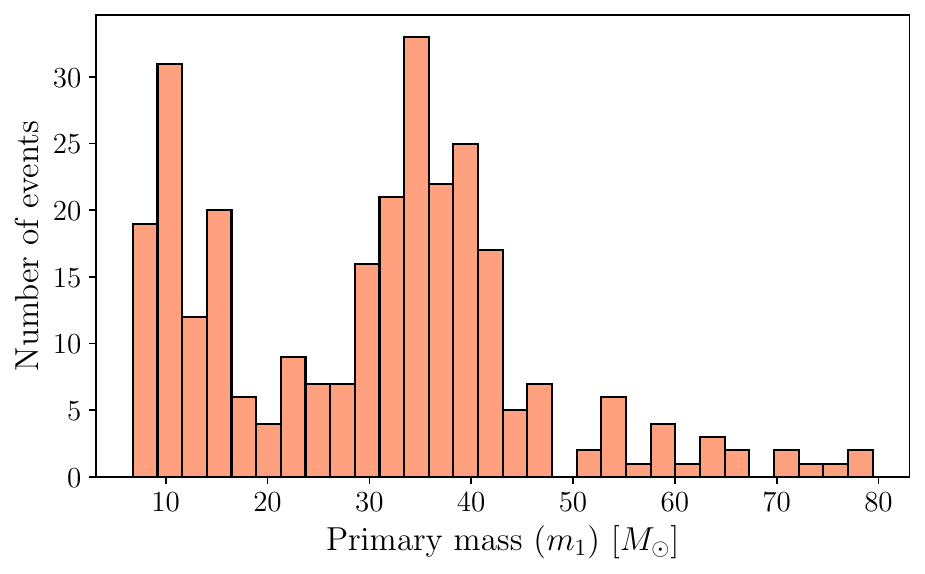}}
   \subfigure[]{\label{fig:Pm2}
    \centering
    \includegraphics[width=\linewidth,trim={0.cm 0  0 0.cm},clip]{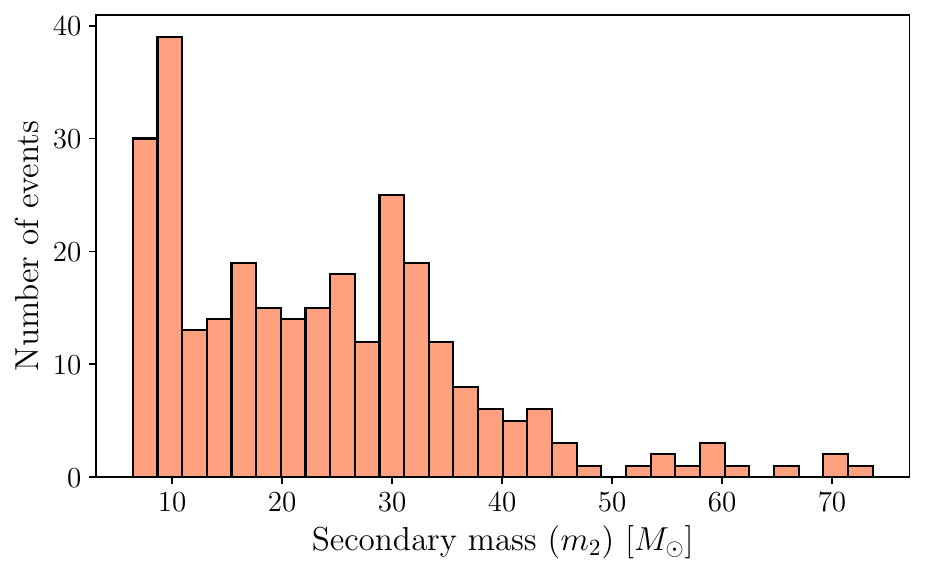}}
  \caption{Histogram of the number of detected events as a function of a) luminosity distance, b) detector-frame primary mass, c) detector-frame secondary mass, for the Vanilla scenario.}
  \label{Hist1}
\end{figure}

We assume black holes to have an astrophysical origin and consequently model the binary merger rate $R(z)$ with a function similar to the one used for fitting the star formation rate (SFR)   \citep{Madau:2014bja}. We use a parametric form for the BBHs merger rate evolution captured by the parameters $\gamma$ and  $\kappa$ as
\begin{equation}
\label{eq:madau}
    R(z) = R_{0} (1+z)^{\gamma} \frac{1+(1+z_p)^{-(\gamma+\kappa)}}{1+(\frac{1+z}{1+z_p})^{(\gamma+\kappa)}},
\end{equation}
where $R_0$ is the merger rate of BBHs at $z-0$ and $z_p$ denotes the peak of the merger rate. The simulated events are drawn from the \textsc{PLG} source frame probability distribution, motivated by the analysis of previously detected GWs events \citep{KAGRA:2021duu, Karathanasis:2022rtr}, which can be analytically expressed by 
\begin{widetext}
\begin{equation}
    P_1(m_1|m_{\rm min},m_{\rm max}, \alpha) =
    \begin{cases}
        \left(1-\lambda\right) {\mathcal{P}}\left(m_1,-\alpha\right) + \lambda ~ G\left(m_1, \mu,\sigma\right), &m_{\rm min} < m_2 < m_{\rm max},\\
        0, &\text{otherwise}
    \end{cases}
\label{Pz1}
\end{equation}
\begin{equation}
   P_2(m_2|m_{\rm min}, m_1, \beta) =
   \begin{cases}
       {\mathcal{P}}\left(m_2,-\beta\right), &m_{\rm min} < m_2 < m_1,\\
       0, &\text{otherwise}
   \end{cases}
\end{equation}
\end{widetext}
where $m_1$ and $m_2$ are source-frame primary and secondary mass respectively. $\mathcal{P}$ and $\textsc{G}$ are normalized power-law function and Gaussian function respectively. For this case 286 GW events are  {detected} with an SNR threshold of $\rho_{\rm th}= 10$ for three years of observation time and duty cycle as given in Tab. \ref{tab:duty}.
The sampled events detector-frame mass distributions are shown in  Figs.~\ref{fig:Pm1}-\ref{fig:Pm2}.
The mock event distribution as a function of distance is shown in Fig.~\ref{fig:Pd1}. The population and rate parameters used to obtain the events are given in Tab.~\ref{tab:injections}.

\begin{figure}
    \centering
    \includegraphics[width=8cm]{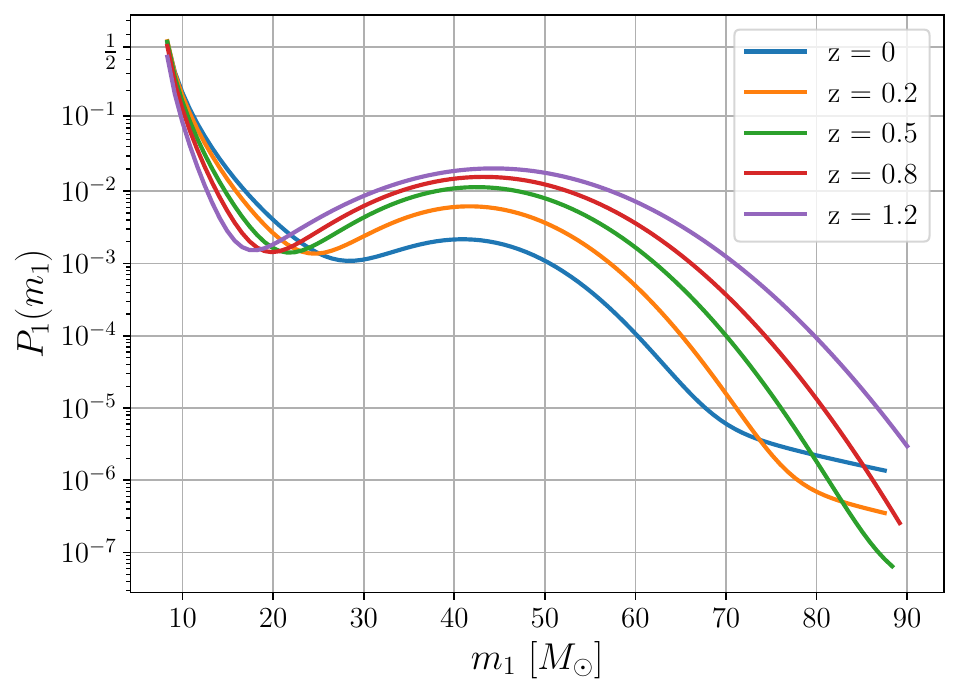}
    \caption{Primary mass distribution for the Redshift-Dependent scenario at different redshifts.}
    \label{fig:Pz1}
\end{figure}

\begin{figure}
    \centering
    \includegraphics[width=8cm]{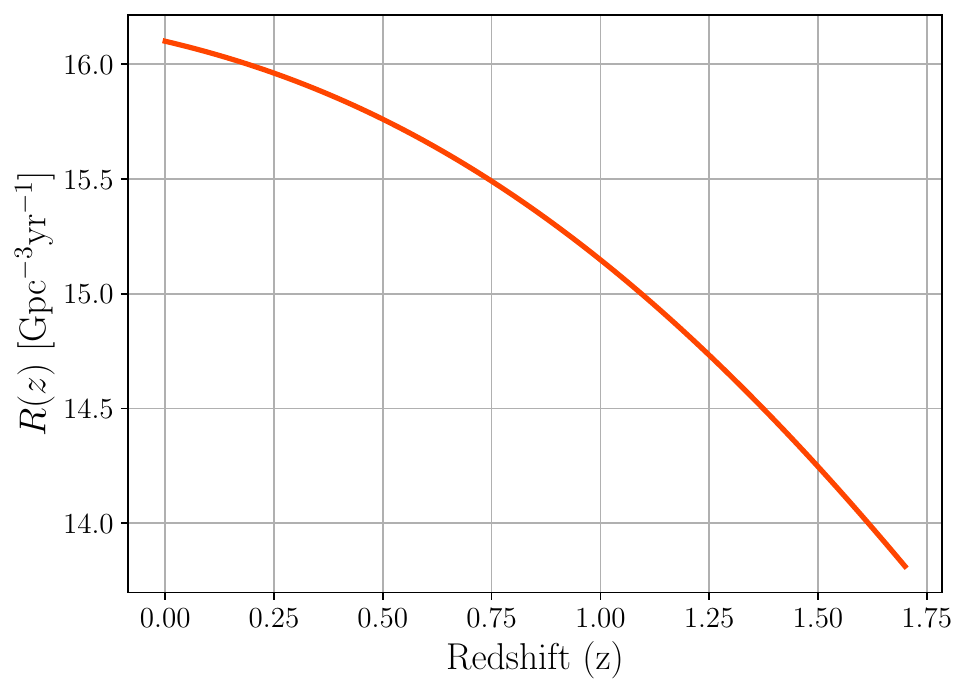}
    \caption{Merger rate as a function of redshift for the Redshift-Dependent scenario with the fiducial values mentioned in Tab.  \ref{tab:injections}.}
    \label{fig:Rz}
\end{figure}

\begin{figure}
  \subfigure[]{\label{fig:dl_2}
    \centering
    \includegraphics[width=\linewidth,trim={0.cm 0  0 0.cm},clip]{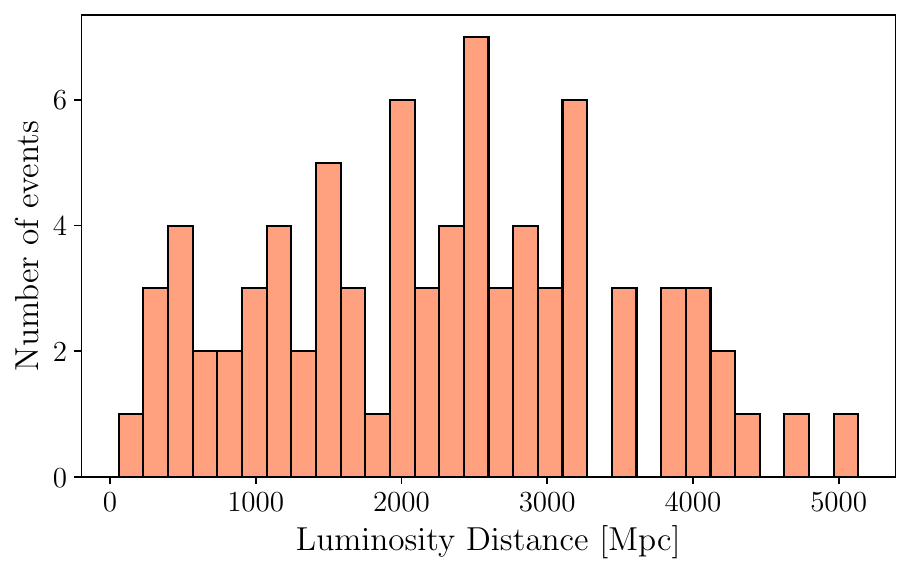}}
  \subfigure[]{\label{fig:m1_2}
    \centering
    \includegraphics[width=\linewidth,trim={0.cm 0cm  0cm 0.cm},clip]{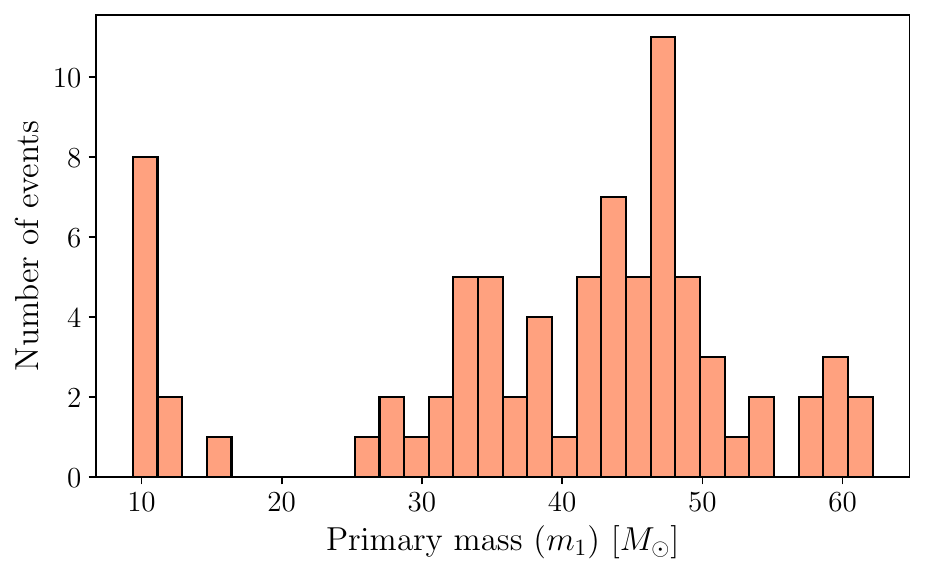}}
   \subfigure[]{\label{fig:m2_2}
    \centering
    \includegraphics[width=\linewidth,trim={0.cm 0  0 0.cm},clip]{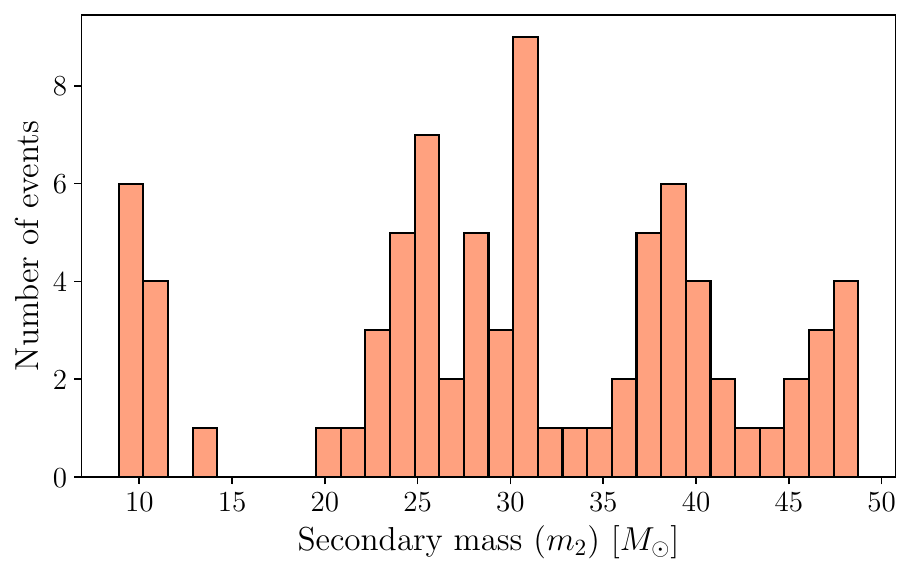}}
  \caption{Histogram of the number of detected events as a function of a) luminosity distance, b) detector-frame primary mass, c) detector-frame secondary mass, for Redshift-Dependent scenario.}
  \label{Hist2}
\end{figure}

\subsection{Redshift-Dependent Scenario}\label{sec:redshift-dependent}
In this scenario, we make the same assumptions as the Vanilla model, but the mass model parameters now include an additional linear dependence on redshift. In this case, we obtained events with an SNR greater than an SNR threshold $\rho_{\rm th}=12$ with one year of observation time and duty cycle as given in Tab. \ref{tab:duty}. The total number of events selected is $80$ for this case. The primary mass and secondary mass distribution is modeled with all parameters of the mass distributions varying with redshift as
\begin{widetext}
\begin{equation}
\label{eq:mass1_redshift_case}
    P_1(m_1|m_{\rm min},m_{\rm max}, \alpha, z) =
    \begin{cases}
        \left((1-\lambda(z)\right) {\mathcal{P}}\left(m_1,-\alpha(z)\right) + \lambda(z) ~ G\left(m_1, \mu(z),\sigma(z)\right), &m_{\rm min}(z) < m_2 < m_{\rm max}(z),\\
        0, &\text{otherwise}
    \end{cases}
\end{equation}
\begin{equation}
   P_2(m_2|m_{\rm min}, m_1, \beta, z) = 
   \begin{cases}
       {\mathcal{P}}\left(m_2,-\beta(z)\right), &m_{\rm min}(z) < m_2 < m_1(z),\\
       0, &\text{otherwise},
   \end{cases}
\end{equation}
\end{widetext}

We model the redshift dependence of each parameter using a linear relation of the form $x(z) = x_0 + z\,\epsilon_x$, where $x_0$ denotes the parameter value at redshift $z = 0$ \citep{Karathanasis:2022hrb}. This form of redshift evolution is applied uniformly across all mass distribution parameters in the redshift-dependent scenario.  {The specific injected values for these parameters are summarized in Table~\ref{tab:injections}, while the prior distributions used for sampling are detailed in Table~\ref{tab:priors_parameter}. The high value of the Hubble constant chosen in the blinded setup checks for the robustness of the analysis and scrutinizes whether there is any sort of confirmation bias in the inference of the Hubble constant. We demonstrate later that this large value of the Hubble constant is not driving any bias in the inference.}

The motivation for introducing redshift dependence arises from the fact that stellar evolution is significantly influenced by the environment in which stars form. Studies show that in low-metallicity environments, more massive stars are more likely to form compared to high-metallicity environments \citep{dopcke2013initial,li2023stellar}. This occurs because low-metallicity gases are less efficient at cooling, leading to a higher Jeans mass and reducing the likelihood of fragmentation into smaller stars \citep{clarke2003characteristic,dopcke2013initial}. Additionally, stars with higher metallicity tend to lose more mass through stellar winds compared to their low-metallicity counterparts, resulting in lighter remnants \citep{vink2001mass,van2005metallicity,mokiem2007empirical}. Consequently, BHs are expected to be more massive at high redshifts, where metallicity is lower compared to low redshifts. The impact of metallicity evolution on BBHs population from GWTC-3 was previously studied,  indicating
 mild hints towards redshift-dependent mass distribution \citep{Mukherjee:2021rtw, Karathanasis:2022rtr}.  {However, no conclusive evidence of it is not found from GWTC-3 \citep{Callister:2024cdx}.}

The primary mass distribution for the Redshift-Dependent scenario is illustrated in Fig. \ref{fig:Pz1}. The peak ($\mu_g$) of the \textsc{Gaussian} component of the primary mass distribution is located at $40.9 \, M_{\odot}$ at $z=0$, shifting to higher masses at a rate of $2.84\,M_{\odot}$ per unit redshift. Similarly, the standard deviation of the \textsc{Gaussian} ($\sigma_g$) is $7.63\,  M_{\odot}$ at $z=0$, increasing linearly with redshift at a rate of $2.7  \,M_{\odot}$. The power-law index of the primary mass distribution takes a value of 5.3 at z = 0, increasing linearly with redshift at a rate of $2.99$, while the power-law index of the secondary mass distribution decreases at a rate of $1.063$ per unit redshift.

For the merger rate, we assume a Madau-Dickinson-like redshift evolution \citep{Madau2014}, similar to that in a redshift-independent vanilla scenario (see Eq. \eqref{eq:madau}). The merger rate for the injected parameters is relatively flat (see Fig. \ref{fig:Rz}), with a maximum value at z = 0 before gradually decreasing with increasing redshift. The maximum redshift of the injection is 1.7.  {
The sources are located at relatively smaller luminosity distances compared to the vanilla scenario, primarily due to the high injected value of $H_0 = 173.28~\rm km\,s^{-1}\,Mpc^{-1}$. The higher detection threshold ($\rho_{\rm th}$) also contributes to selecting nearer sources, but its impact on the overall distance distribution is subdominant. It is important to note that the redshift dependence in the simulated mass distribution is implemented directly in redshift space. Therefore, the compression in the luminosity distance distribution caused by the high $H_0$ does not influence the effect of the redshift evolution in the mass distribution.
}

In Figs. \ref{fig:dl_2},\ref{fig:m1_2}, and \ref{fig:m2_2}, we depict the histogram of luminosity distance ($d_{L}$) and the component masses ($m_{1}$, and $m_{2}$) respectively of the injected sources.  The minimum mass is set to $7.8 M_{\odot}$ at $z = 0$, with only a tiny redshift dependence. Compared to the vanilla scenario, there is a relatively smaller fraction of sources between $10 \, M_{\odot}$ and $30 \, M_{\odot}$, which is a consequence of the steeper power law for both the primary and secondary mass distributions. The luminosity distance of the sources in this scenario is more concentrated at lower values compared to the vanilla scenario. This is because the merger rate here is relatively constant with redshift, whereas, in the vanilla scenario, it increases steeply with redshift.

\begin{table*}
\caption{Summary of injected values for the two scenarios: Vanilla and Redshift-Dependent. We report a description for each of the parameters in the cosmological assumptions ($\Lambda$CDM) with different $H_0$ values, the mass model (\textsc{Power Law + Gaussian Peak}, with redshift-dependence in the second scenario), and the Madau-Dickinson rate model with different parameters.}
\label{tab:injections}
\centering
\begin{tabular}{l l r r}
& &\multicolumn{2}{c}{\bf Injected Value} \\
\bf Parameter &\bf Description &\bf Vanilla &\bf Redshift-Dependent\\
\hline
\hline
&\texttt{$\Lambda$CDM Cosmological model} &&\\
\hline
\hline
$H_0$ &Hubble constant in [km s$^{-1}$Mpc$^{-1}$] &$67.8$ &$173.28$\\
$\Omega_{\rm m, 0}$ &Matter energy density today &$0.3$ &$0.3$\\
\hline
\hline
&\texttt{Power Law + Gaussian Peak mass model} &&\\
\hline
\hline
$\alpha$ &PL index of primary mass &$3.4$ &$5.325+2.99z$\\
$\beta$ &PL index of secondary mass &$0.8$ &$3.05-1.063z$\\
$m_{\rm min}$ &minimum source mass in $M_{\odot}$ &$5$ &$7.8+0.324z$\\
$m_{\rm max}$ &maximum source mass in $M_{\odot}$ &$100$ &$87.58+ 17.16z$\\
$\delta_m$ &smoothing factor in $M_{\odot}$ at low-mass cut-off &$4.8$ &$3.19$\\
$\mu_g$ &peak of the Gaussian in $M_{\odot}$ &$35$ &$40.9+2.84z$\\
$\sigma_g$ &sigma of the Gaussian in $M_{\odot}$ &$3.9$ &$7.63+2.70z$\\
$\lambda_{\rm peak}$ &fraction of events in Gaussian in $[0,1]$ interval &$0.04$ &$0.04+0.43z$\\
\hline
\hline
&\texttt{Madau-Dickinson rate model} &&\\
\hline
\hline
$\gamma$    &Power law exponent of rate ($z\lesssim z_{\rm p}$)  &$2.7$ &$0.03$\\
$\kappa$    &(Negative of) PL exponent of rate ($z\gtrsim z_{\rm p}$)   &$2.9$ &$2.92$\\
$z_{\rm p}$     &Rate parameter (turnover point)   &$1.9$ &$3.86$\\
$R_0$   &Local merger rate in [Gpc$^{-3}$yr$^{-1}$]   &$20$ &$16.1$\\
\hline
\label{Table1}
\end{tabular}
\end{table*}

\begin{table*}
\caption{Prior ranges used to sample the injection parameters for the redshift-dependent scenario.}
\label{tab:priors_parameter}
\centering
\begin{tabular}{l r}
\textbf{Parameter} & \textbf{Prior used for sampling} \\
\hline
\hline
$H_0$ & $\mathcal{U}(10, 200)$ \\
$\Omega_{\rm m, 0}$ & $0.3$ (Fixed) \\
\hline
\hline
$\alpha_0$ & $\mathcal{U}(1.5, 12)$ \\
$\beta_0$ & $\mathcal{U}(-4, 12)$ \\
$m_{\rm min}$ & $\mathcal{U}(2, 10)$ \\
$m_{\rm max}$ & $\mathcal{U}(70, 200)$ \\
$\delta_m$ & $\mathcal{U}(0, 10)$ \\
$\mu_g$ & $\mathcal{U}(10, 70)$ \\
$\sigma_g$ & $\mathcal{U}(0.4, 10)$ \\
$\lambda_{\rm peak}$ & $\mathcal{U}(0, 1)$ \\
\hline
\hline
$\epsilon_\alpha$ & $\mathcal{U}\left(-\frac{\alpha}{z_{\rm max}}, \frac{\alpha}{z_{\rm max}}\right)$ \\
$\epsilon_\beta$ & $\mathcal{U}\left(-\frac{\beta}{z_{\rm max}}, \frac{\beta}{z_{\rm max}}\right)$ \\
$\epsilon_{m_{\rm min}}$ & $\mathcal{U}\left(0, \frac{m_{\rm min}}{z_{\rm max}}\right)$ \\
$\epsilon_{m_{\rm max}}$ & $\mathcal{U}\left(0, \frac{m_{\rm max}}{z_{\rm max}}\right)$ \\
$\epsilon_{\mu_g}$ & $\mathcal{U}\left(\max\left(0, \epsilon_{m_{\rm min}} + \frac{m_{\rm min} - \mu_g}{z_{\rm max}}\right), \min\left(\frac{\mu_g}{z_{\rm max}}, \epsilon_{m_{\rm max}} + \frac{m_{\rm max} - \mu_g}{z_{\rm max}}\right)\right)$ \\
$\epsilon_{\sigma_g}$ & $\mathcal{U}\left(0, \frac{\sigma_g}{z_{\rm max}}\right)$ \\
$\epsilon_\lambda$ & $\mathcal{U}\left(-\frac{\lambda}{z_{\rm max}}, \frac{1 - \lambda}{z_{\rm max}}\right)$ \\
\hline
\hline
$\gamma$ & $\mathcal{U}(0, 12)$ \\
$\kappa$ & $\mathcal{U}(0, 6)$ \\
$z_{\rm p}$ & $\mathcal{U}(0, 4)$ \\
$R_0$ & $\mathcal{U}(0, 100)$ \\
\hline
\end{tabular}
\end{table*}

\section{Methods}\label{sec-formalism}
In what follows, we first give an overview of the spectral sirens method used to infer at the same time cosmological and population parameters given the two starting sets of BBH events: one for the Vanilla case and one for the Redshift-Dependent scenario. We present a reconstruction of both the rate and mass models for two cases, comparing them to the distributions used to sample the two populations. The main results on the inference of $H_0$ are then presented and discussed in full details\footnote{All the scripts used to produce the results in this paper are available on GitLab \href{https://git.ligo.org/konstantin.leyde/cosmology_mdc_analysis_group_1_summary/-/tree/main/}{repository}.}.

\subsection{Spectral sirens method: \texttt{icarogw} setup}
We use the \icarogw{} package \citep{Mastrogiovanni:2023zbw} to infer the population and cosmological parameters simultaneously (a brief description of all the parameters is provided in Tab.~\ref{tab:injections}).
Within this package, we use \bilby{}'s \citep{bilby_paper} sampler \texttt{dynesty} \citep{dynesty_paper} to generate posterior samples for the hyperparameters. 
The source distributions assumed in the analysis coincide with the ones used during the generation of the GW population only in the Vanilla case. For the Redshift-Dependent scenario, we keep the same assumptions of the Vanilla analysis, even though the underlying population is now Redshift-Dependent.

The combined posteriors on the source population and cosmological hyper-parameters, given $N_{\rm obs}$ GW detections each with data $\{x\}=(x_1, x_2, \dots, x_{\rm obs})$, are obtained using the following equation (\cite{Mandel:2018mve, Vitale:2020aaz, Mastrogiovanni:2021wsd}):
\begin{equation}
    p(\Lambda |\{x\}, N_{\rm obs})\propto \pi (\Lambda)\prod_{i=1}^{N_{\rm obs}} \frac{\int p(x_i|\Lambda, \theta)p_{\rm pop}(\theta|\Lambda)d\theta}{\int p_{\rm det}(\theta,\Lambda)p_{\rm pop}(\theta|\Lambda)d\theta},
\label{eq:combined_posterior}
\end{equation}
where $\pi(\Lambda)$ is the prior on the hyper-parameters (both population parameters and cosmological ones, see Tab.~\ref{tab:priors}). Since each GW event is independent, we multiply across the $N_{\rm obs}$ observations. The probability of each GW event can be broken down into two main components: the parameters of the individual source, represented by $\theta$, and the population-induced prior, $p_{\rm pop}$, which describes the expected distribution of the hyperparameters at the population level.

The denominator of the likelihood, $\int p_{\rm det}p_{\rm pop}\text{d}\theta$, corrects for the selection bias. The data set that is analyzed has been obtained after applying selection criteria. In this study, we required that the events' match-filtering SNRs $\rho$ are above a fixed threshold ($\rho_\text{th}=10$ for the Vanilla case and $\rho_\text{th}=12$ for the Redshift-Dependent scenario). This selection must be accounted for, to avoid the Malmquist bias \citep{1922MeLuF.100....1M, 2004AIPC..735..195L, Mandel:2018mve}. The correction method consists of estimating the probability of detection of gravitational wave events, using the very same selection criterium and for any values of hyperparameters $\Lambda$, describing the cosmology, the population parameters, and the mergers rate (see Tab. ~\ref{tab:injections}). This probability of detection is given by the ratio of the expected number of detected events $N_\mathrm{exp}(\Lambda)$ to the total number of mergers $N(\Lambda)$:
\begin{equation}
\frac{N_\mathrm{exp}(\Lambda)}{N(\Lambda)}=\int p_{\rm det}(\theta,\Lambda)p_{\rm pop}(\theta|\Lambda)\text{d}\theta,
\end{equation}
$\theta$ being the set of the individual parameters of the binary system: $\theta=(m_{1}, m_{2}, z, \iota, \mathrm{RA}, \mathrm{Dec}...)$. We chose to study a population model that only depends on the mass and redshift so that we restrict $\theta$ to $\theta=(m_{1}, m_{2}, z)$ and we consider the domain where the events are detected, i.e. when $p_\text{det}=1$:
\begin{equation}
\frac{N_\mathrm{exp}(\Lambda)}{N(\Lambda)}=\int_{\rho\geqslant \rho_\text{th}}p_{\rm pop}(m_{1s}, m_{2s}, z|\Lambda)\text{d}m_{1}\text{d}m_{2}\text{d}{z}.
\end{equation}
The last integral is estimated by Monte-Carlo integration using a large set of $N_\text{sim}$  {realizations} $(m_{1,i},m_{2,i},z_i)$ corresponding to $N_\text{det}$ detected ($\rho\geqslant \rho_\text{th}$) simulated events. These injected events are randomly drawn from an initial probability density function $\pi(m_{1}, m_{2}, z|\Lambda)$ so that the probability of detection is evaluated as:
\begin{align}
\frac{N_\mathrm{exp}(\Lambda)}{N(\Lambda)}&=\int_{\rho\geqslant \rho_\text{th}}p_{\rm pop}(m_{1}, m_{2}, z|\Lambda)\text{d}m_{1}\text{d}m_{2}\text{d}{z},\nonumber\\
&=\frac{1}{N_\text{sim}}\sum_{i=1}^{N_\text{det}}\frac{p(m_{1,i},m_{2,i},z_i|\Lambda)}{\pi(m_{1,i},m_{2,i},z_i|\Lambda)}.
\end{align}

\subsection{Priors settings}
We examine $14$ hyperparameters listed in Tab.~\ref{tab:injections}. We assign broad and uninformative priors to each of them. The priors for the hyperparameters in the two scenarios are summarized in Tab.~\ref{tab:priors}. Three parameters  {prior ranges} differ between the two scenarios: $H_0$, $\mu_g$, and $\sigma_g$. Initially, the prior ranges for both scenarios were the same. After our preliminary results exhibited railing on either side of their prior range, we decided to enlarge the prior settings for the Redshift-Dependent scenario.  {We refer to this aspect of the study as the impact from astrophysical prior on $H_0$ inference in the remaining paper. Due to the blinded nature of the MDC (as discussed in Sec. \ref{sec-mdc}), different prior choices are made for the analysis.}  

\begin{table}
\caption{List of priors used in the analysis for both the Vanilla and the Redshift-Dependent scenarios.}
\label{tab:priors}
\centering
\begin{tabular}{l r r}
&\multicolumn{2}{c}{\bf Priors}\\
\hline
\hline
&\bf Vanilla &\bf Redshift-Dependent\\
\hline
\hline
$H_0$ &$\mathcal{U}(30, 140)$ &$\mathcal{U}(10, 250)$\\
$\Omega_{\rm m, 0}$ &$\mathcal{U}(0.1, 0.9)$ &$\mathcal{U}(0.1, 0.9)$\\
\hline
\hline
$\alpha$    &$\mathcal{U}(1.5, 12)$ &$\mathcal{U}(1.5, 12)$\\
$\beta$ &$\mathcal{U}(-4, 12)$ &$\mathcal{U}(-4, 12)$\\
$m_{\rm min}$ &$\mathcal{U}(2, 10)$ &$\mathcal{U}(2, 10)$\\
$m_{\rm max}$ &$\mathcal{U}(50, 200)$ &$\mathcal{U}(50, 200)$\\
$\delta_{\rm m}$ &$\mathcal{U}(0, 10)$ &$\mathcal{U}(0, 10)$\\
$\mu_{\rm g}$ &$\mathcal{U}(20, 50)$ &$\mathcal{U}(10, 80)$\\
$\sigma_{\rm g}$  &$\mathcal{U}(0.4, 10)$  &$\mathcal{U}(0.4, 20)$\\
$\lambda_{\rm peak}$ &$\mathcal{U}(0, 1)$ &$\mathcal{U}(0, 1)$\\
\hline
\hline
$\gamma$ &$\mathcal{U}(0, 12)$ &$\mathcal{U}(0, 12)$\\
$\kappa$  &$\mathcal{U}(0, 6)$ &$\mathcal{U}(0, 6)$\\
$z_{\rm p}$ &$\mathcal{U}(0, 4)$ &$\mathcal{U}(0, 4)$\\
$R_0$ &$\mathcal{\log U}(10^{-2}, 10^3)$ &$\mathcal{\log U}(10^{-2}, 10^3)$\\
\hline
\end{tabular}
\end{table}

\section{Results from the \texttt{Blinded-MDC}}\label{sec-results}

\subsection{Vanilla model results}
\begin{figure}[]
    \centering
    \includegraphics[scale = 0.55]{ 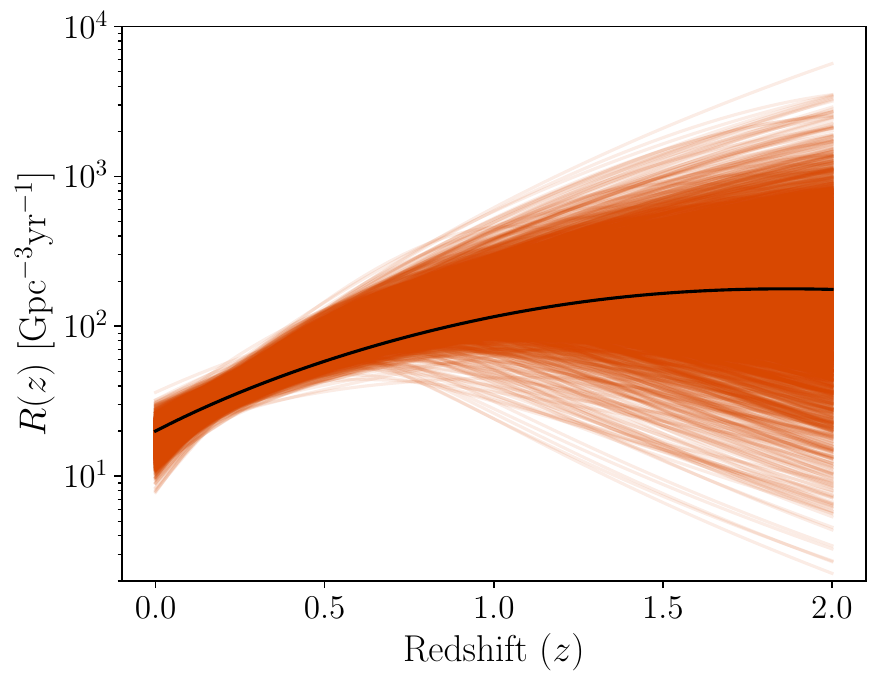}
    \caption{Posterior samples plots for the Vanilla scenario following Eq.~\ref{eq:madau}. The black line represents the injected values. The colored lines show the posterior samples from the $68\%$ highest density interval.}
    \label{fig:vanilla_rate_samples}
\end{figure}

\begin{figure}[]
    \centering
    \includegraphics[scale = 0.55]{ 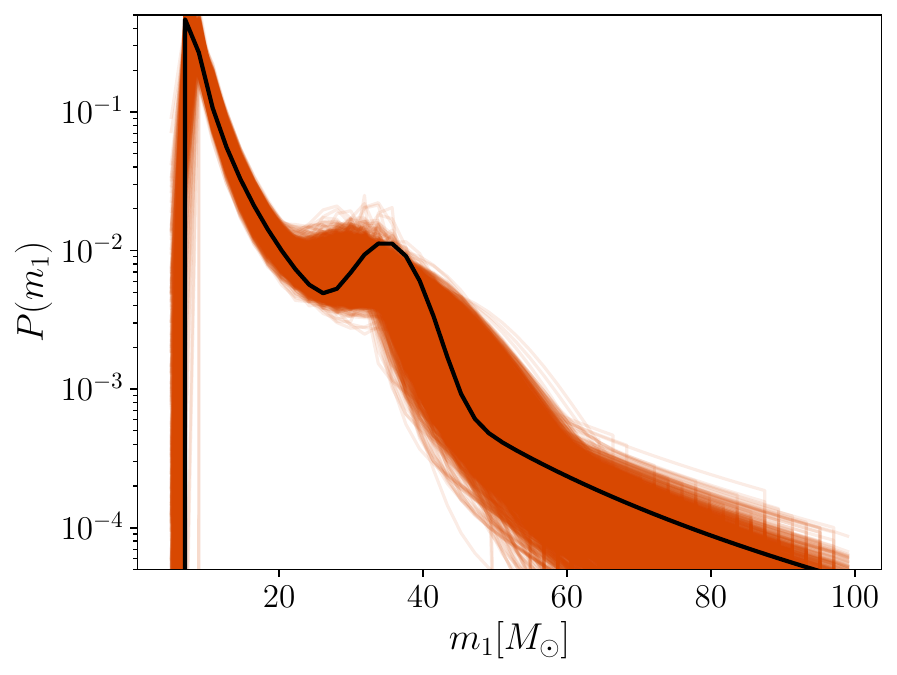}
    \caption{Posterior samples plots for the Vanilla scenario for the primary mass distribution following Eq.~\ref{Pz1}. The black line represents the injected values. The colored lines show the posterior samples from the $68\%$ highest density interval.}
    \label{fig:vanilla_mass_samples}
\end{figure}

 {The results of the Vanilla model serve as an overall validation of the setup and demonstrate its consistency with the initial assumptions. We start from the reconstruction of both the rate and mass models, as shown in Figs.~\ref{fig:vanilla_rate_samples} and~\ref{fig:vanilla_mass_samples}. The colored lines represent the posterior samples from the $68\%$ highest density interval, while the solid black lines indicate the starting distribution derived from the injection values in Tab.~\ref{tab:injections}. We find good agreement with the injected rate and mass distributions.}

\begin{figure}[]
    \centering
    \includegraphics[scale = 0.425]{ 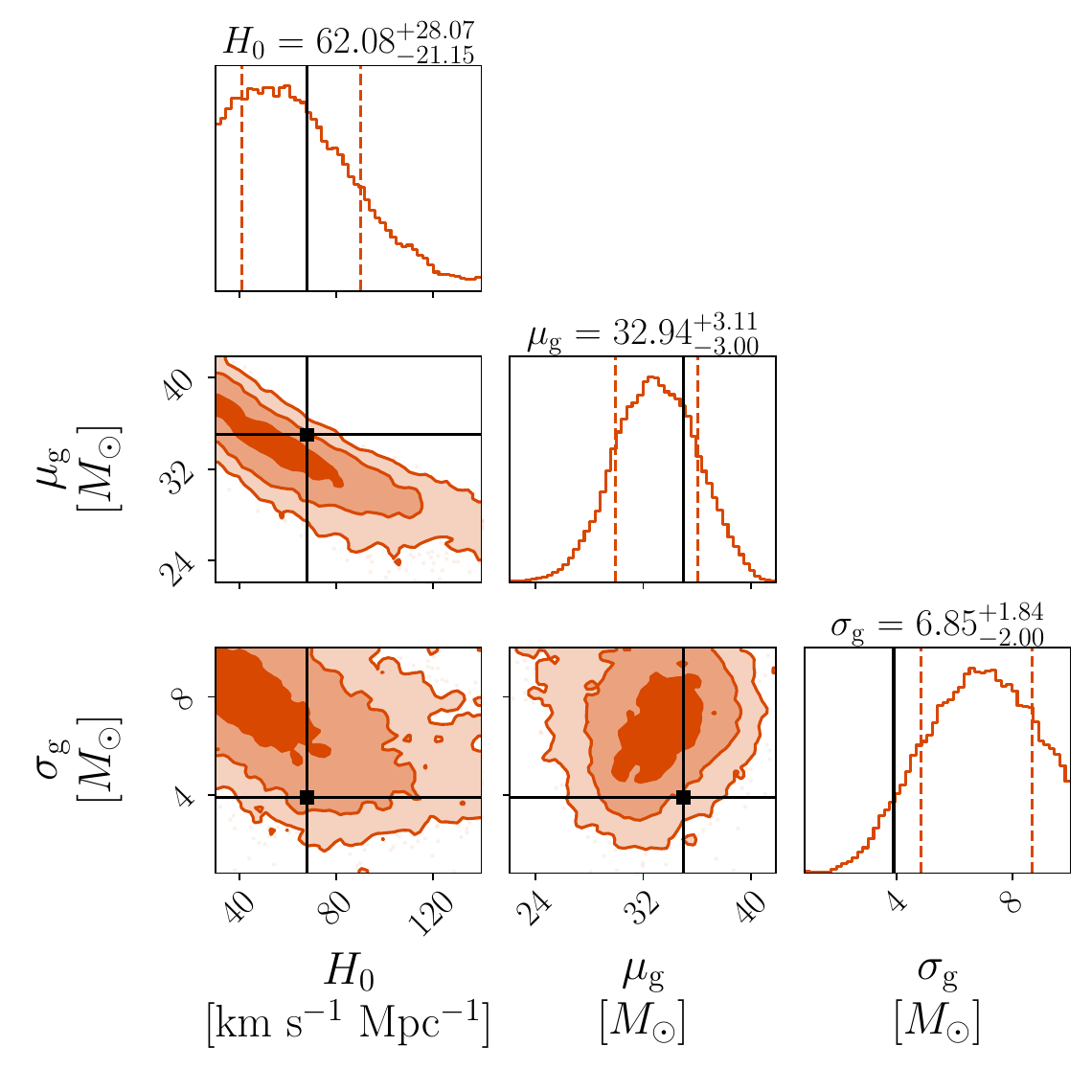}
    \caption{Reduced corner plot for the Vanilla scenario of $H_0$ and parameters that mostly correlate with H$_0$, which are $\mu_g$ and $\sigma_g$. The injected (true) values are plotted as black lines. The three different colored regions of the 2D posterior represent, from the darkest to the lighter shade, respectively, the $39.3\%$, $86.5\%$, and $98.9\%$\,C.L. contours for a 2D Gaussian distribution. The dashed lines in the corresponding 1D histograms instead show the $68\%$\,C.L. of the marginalized distribution. For the full corner plot refer to Fig. \ref{fig:vanilla_corner_full}.}
    \label{fig:vanilla_red_corner}
\end{figure}

 {Our estimate for $H_0$ using all 286 events and the full set of $10^6$ injections is $62.08^{+28.07}_{-21.15}$ km s$^{-1}$Mpc$^{-1}$ at $1\sigma$, compared to the injected value of $67.8$ km s$^{-1}$Mpc$^{-1}$. In most cases, the inferred values of parameters are recovered at the $1\sigma$ level, with some exceptions like $\alpha$, $m_{\rm min}$, $\sigma_g$, and $\gamma$, which are slightly outside the $1\sigma$ range but within $2\sigma$, as can be seen from the full corner plot in Fig.~\ref{fig:vanilla_corner_full}. Fig.~\ref{fig:vanilla_red_corner} illustrates a reduced corner plot that highlights the distributions of the $H_0$ posterior samples along with the parameters that demonstrate the strongest correlations with $H_0$, specifically $\mu_g$ and $\sigma_g$, which are negatively correlated with $H_0$. This means that higher values of $\mu_g$ suggest a more massive BBH population, leading to lower inferred redshifts and a lower $H_0$ to match observed signals at specific luminosity distances.}

 {The three different colored regions of the 2D posterior in all the corner plots shown in this paper represent, from the darkest to the lighter shade, respectively, the $39.3\%$, $86.5\%$, and $98.9\%$\,C.L. contours for a 2D Gaussian distribution. The dashed lines in the corresponding 1D histograms instead show the $68\%$\,C.L. of the marginalized distribution.}

\begin{figure}[]
    \centering
    \includegraphics[scale = 0.5]{ 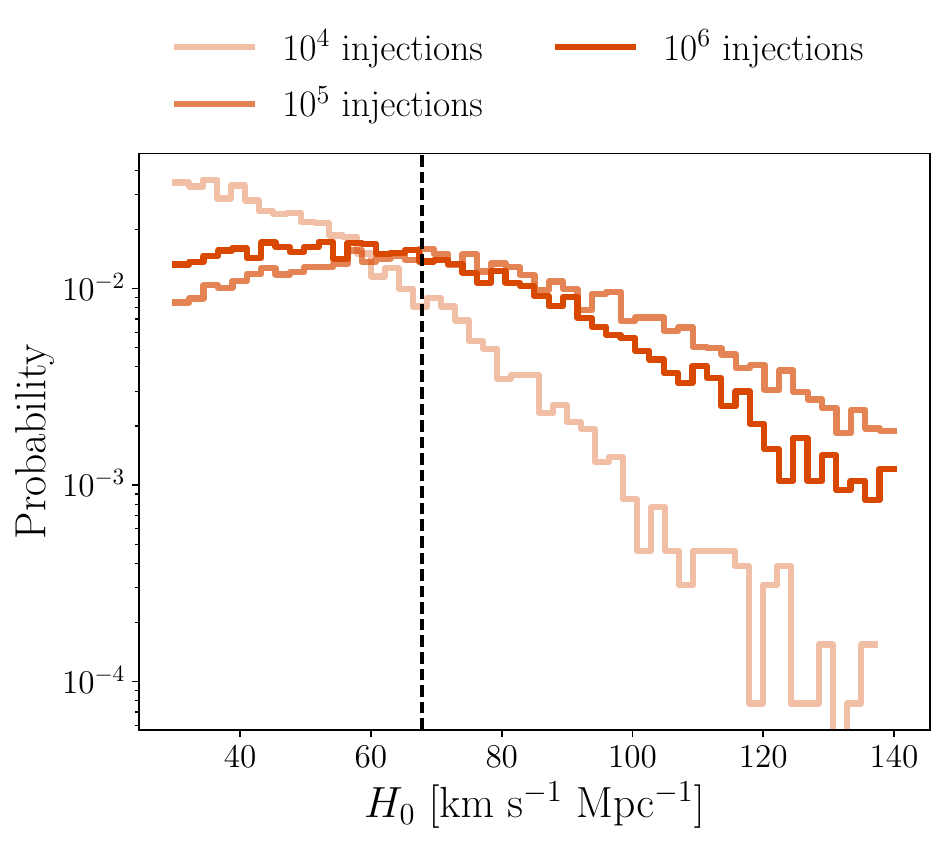}
    \caption{Check with the Vanilla scenario events. Posterior samples of the $H_0$ parameter varying the number of injections. The dashed black line is the injected value. See also the full plot with all the parameters in Fig.~\ref{fig:vanilla_inj_set}).}
    \label{fig:vanilla_inj_set}
\end{figure}

\begin{figure}[]
    \centering
    \includegraphics[scale = 0.5]{ 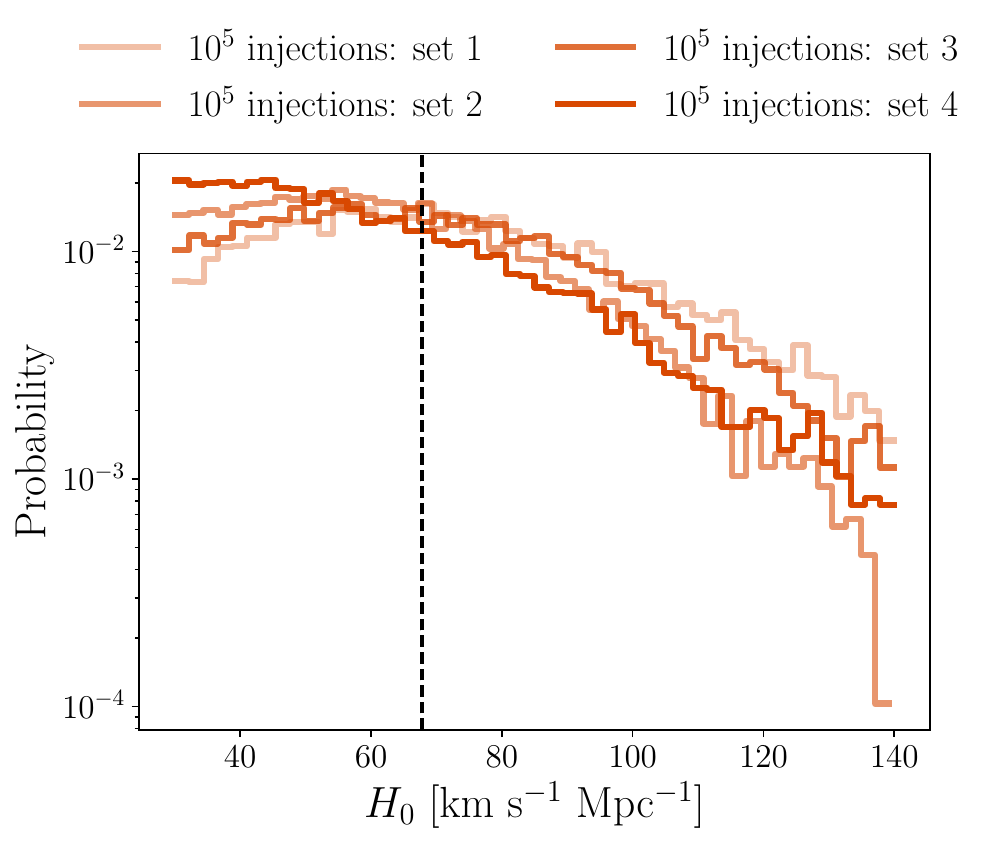}
    \caption{ {Check with the Vanilla scenario events. Posterior samples of the $H_0$ parameter varying the realization sample with $10^5$ injections. The dashed black line is the injected value. See also the full plot with all the parameters in Fig.~\ref{fig:vanilla_full_posteriors_inj_sample}).}}
    \label{fig:vanilla_inj_sample}
\end{figure}

\begin{figure}[]
    \centering
    \includegraphics[scale = 0.5]{ 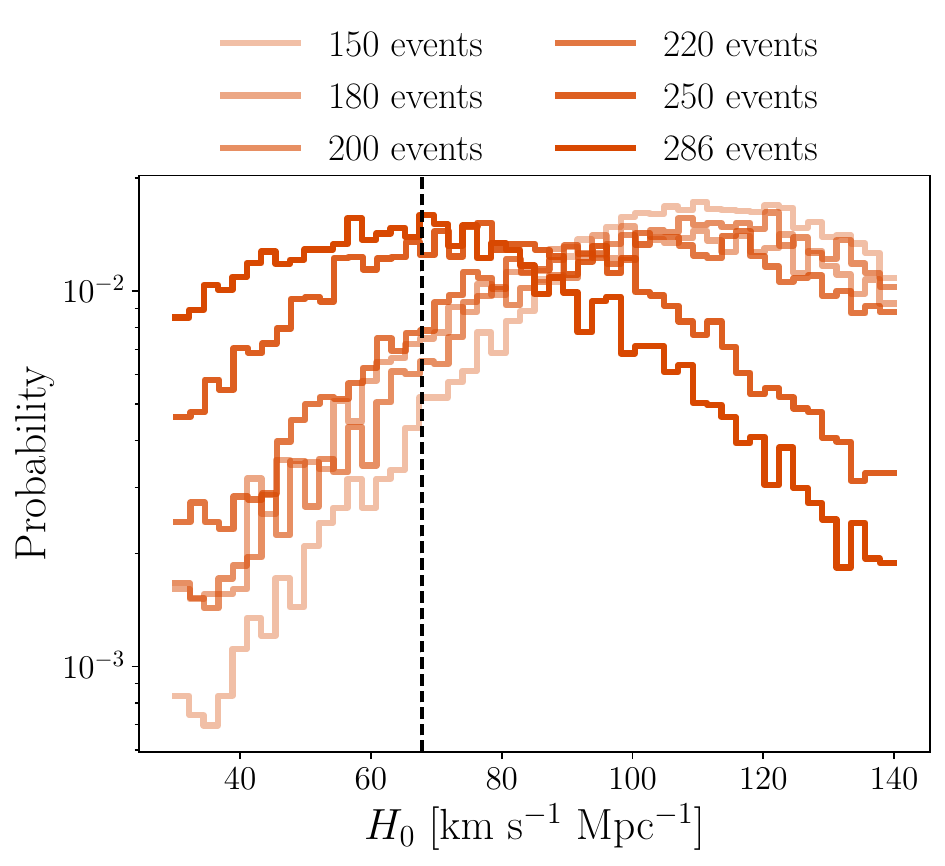}
    \caption{Check with the Vanilla scenario events. Posterior samples for the $H_0$ parameter using a different number of events for the Vanilla case. The dashed black line is the injected value. See also the full plot with all the parameters in Fig.~\ref{fig:vanilla_full_posteriors_ev_num}).}
    \label{fig:vanilla_ev_num}
\end{figure}

\begin{figure}[]
    \centering
    \includegraphics[scale = 0.5]{ 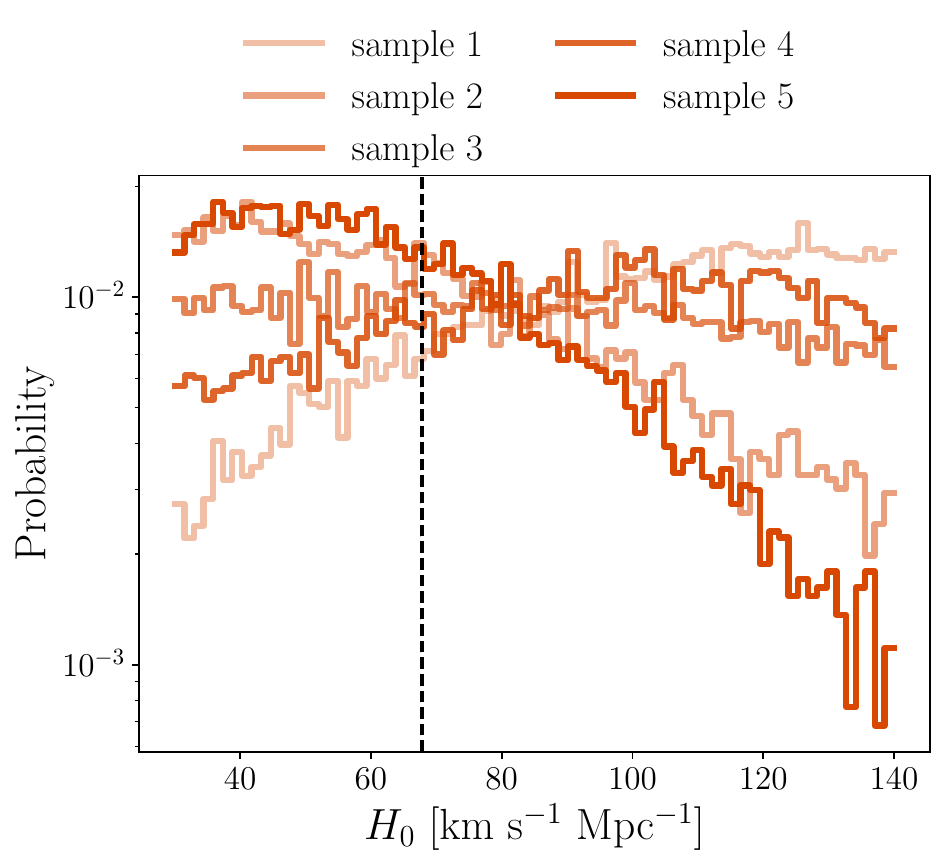}
    \caption{Check with the Vanilla scenario events. Posterior samples of the $H_0$ parameter varying the events sample. The dashed black line is the injected value. See also the full plot with all the parameters in Fig.~\ref{fig:vanilla_full_posteriors_pe_sample}).}
    \label{fig:vanilla_pe_set}
\end{figure}

We then performed the following series of checks to pinpoint potential sources of statistical uncertainties:
\begin{itemize}
 \item {\bf Impact of the injection samples: number and realization}: to study the robustness of the population inference scheme, we vary the realization of GW signals, (referred to as injection samples), that are used to compute the denominator of Eq.~\eqref{eq:combined_posterior}. 
    The denominator is evaluated as a Monte-Carlo sum, hence we expect to have more accurate results for a larger amount of samples. 
    The hyper-parameter posterior is obtained for three simulations, with $10^4,\, 10^5$, and $10^6$ samples. We generally find that the inferred values of all hyper-parameters are compatible with their true values at $90\%$ credible region. In Fig.~\ref{fig:vanilla_inj_set}, we show $H_0$ posteriors for the three cases. We also varied the sample of $10^5$ injections and obtained comparable results, as shown in Fig.~\ref{fig:vanilla_inj_sample}. The distribution 
    confirms that the evaluation of the denominator partially depends on the number of injections and not the specific realization of the sample, {once we reached a sufficient number of injections for the denominator to converge, which in our case means $10^5$ injections.}. The posteriors for all the parameters are shown in appendix~\ref{app: additional results vanilla model} in Fig.~\ref{fig:vanilla_full_posteriors_inj_set} varying the number of injections and in Fig.~\ref{fig:vanilla_full_posteriors_inj_sample} varying the injections sample. 
    \item {\bf Dependence on the number of events}: we study how the inferred cosmological and population parameters depend on the number of GW events used for the analysis. We compare the results using different numbers of GW events, specifically $[150, 180, 200, 220, 250, 286]$, with $10^5$ injections (once we assessed $10^5$ and $10^6$ injections were producing the same outcome, as shown in the previous point). The results shown in Fig.~\ref{fig:vanilla_ev_num} demonstrate that we obtain more accurate constraints on the parameters as more GW events are used. In Fig.~\ref{fig:vanilla_ev_num} we show $H_0$ posteriors only for clarity (the full parameter set is displayed in appendix~\ref{app: additional results vanilla model} in Fig.~\ref{fig:vanilla_full_posteriors_ev_num}).
    \item {\bf Dependence on event sample realization}: due to the homogeneity principle, the analysis of subsets of the observed gravitational wave catalog is expected to yield parameter values with random deviations (which statistically average out) from the parameter values extracted from the whole catalog. For further validation of the analysis scripts, we test this expectation. We split the observed GW catalog in batches of 50 and analyze the resulting sub-populations with $10^5$ injections used to compute the denominator of Eq.~\eqref{eq:combined_posterior}. 
    In Fig.~\ref{fig:vanilla_pe_set} we show some example results for the posterior distribution restricted to $H_0$. We find that there are no outliers in our sets of events. The posteriors are summarized in appendix~\ref{app: additional results vanilla model} in Fig.~\ref{fig:vanilla_full_posteriors_pe_sample}. 
\end{itemize}

In summary, the \texttt{Blinded-MDC} analysis for the Vanilla scenario shows that when the underlying true model of the astrophysical population is the same as the model used in the analysis, the inference of parameters is not biased. However, due to the degeneracy between the cosmological parameters and the astrophysical population parameters, the inferred value of $H_0$ can shift away from the true value.  {We also show the impact of prior on the astrophysical parameters in inferring $H_0$ in the repository. The plot  {in Fig.~\ref{fig:red_vanilla_corner_larger_priors}} indicates that wider prior on the parameters related to black hole mass distribution moves the $H_0$ posterior to a lower value.}  {The Vanilla model analysis served as a validation tool to assess the robustness of the inference pipeline against statistical and sampling uncertainties under the scenarios analyzed above: injection sample size and realization, number of events, and event sample realization).}

\subsection{Redshift-Dependent Scenario Results}
The goal of analyzing the Redshift-Dependent scenario is to understand the impact of the redshift evolution of the mass distribution on the inference of the Hubble constant due to mis-modeling the astrophysical population model of BBHs by assuming a redshift-independent scenario. So, we use the redshift independent \textsc{PLG} model, which is currently a setup for the population inference, for the analysis of this mock simulations. However, it is important to stress here, that though the analysis is performed for a specific mass model, the impact on the inference of $H_0$ can happen for any other astrophysical population model as well, if there is a mismatch with the simulation. 

\begin{figure}[]
    \centering
    \includegraphics[scale = 0.425]{ 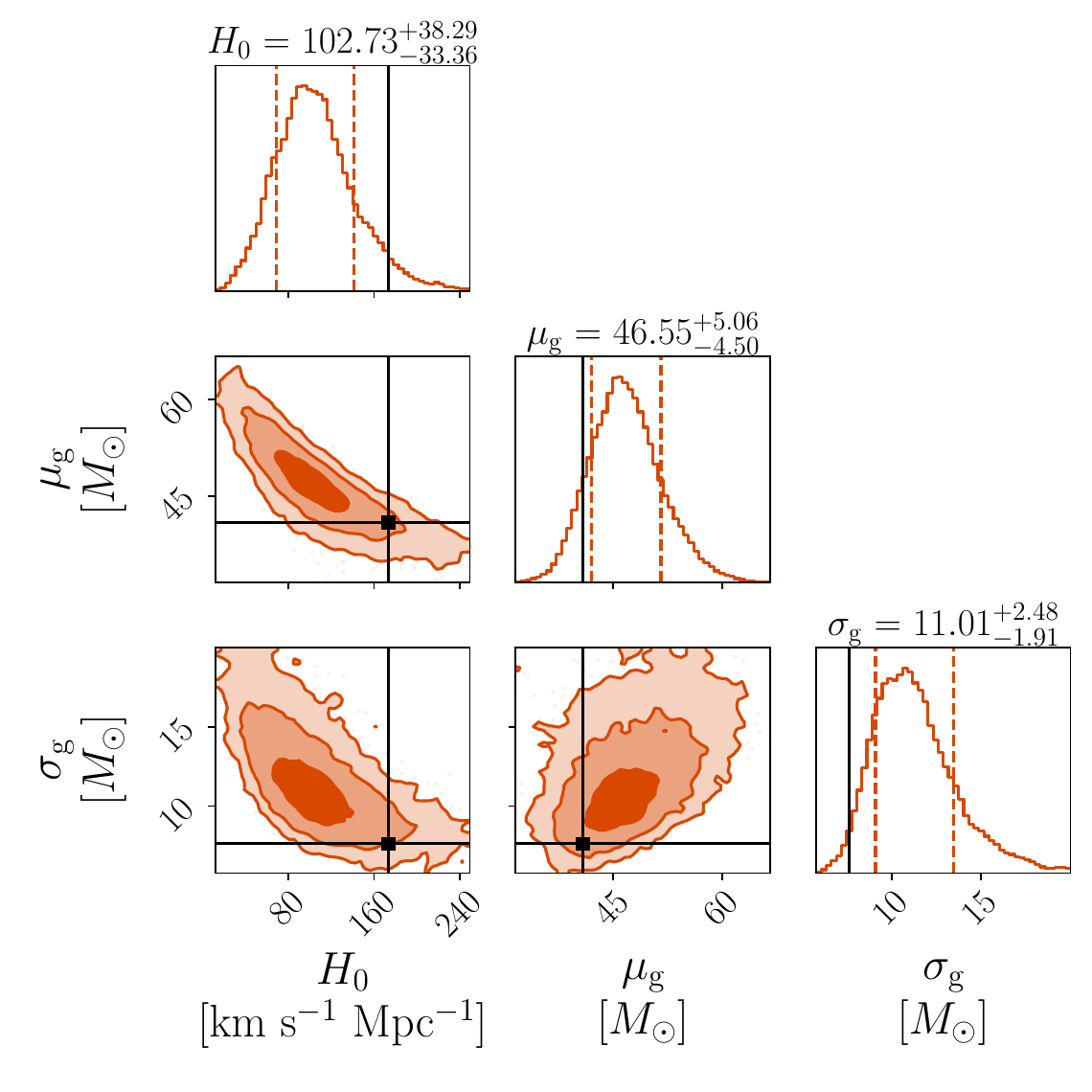}
    \caption{Reduced corner plot for the Redshift-Dependent scenario of $H_0$ and parameters that mostly correlate with H$_0$, which are $\mu_g$ and $\sigma_g$. The injected (true) values are plotted as black lines. The three different colored regions of the 2D posterior represent, from the darkest to the lighter shade, respectively, the $39.3\%$, $86.5\%$, and $98.9\%$\,C.L. contours for a 2D Gaussian distribution. The dashed lines in the corresponding 1D histograms instead show the $68\%$\,C.L. of the marginalized distribution. For the full corner plot refer to Fig. \ref{fig:redshift_corner_full}. See also Fig.~\ref{fig:mu_g} and Fig.~\ref{fig:sigma_g} for a redshift evolution of $\mu_g$ and $\sigma_g$.}
    \label{fig:redshift_red_corner}
\end{figure}

\begin{figure}[]
    \centering
    \includegraphics[scale = 0.5]{ 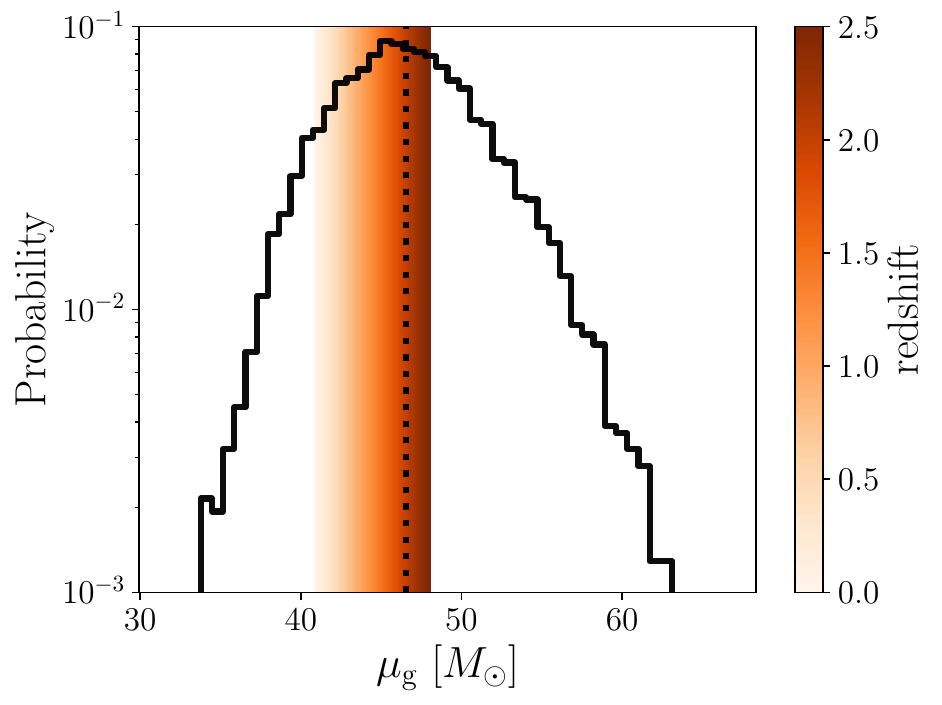}
    \caption{Posterior samples for $\mu_g$ for the Redshift-Dependent scenario are plotted in \textit{black}. The dotted vertical line represents the median of the distribution. The colored region represents the values of the injected $\mu_g$, $40.9+2.84z$, at different redshifts specified in the color bar.}
    \label{fig:mu_g}
\end{figure}

\begin{figure}[]
    \centering
    \includegraphics[scale = 0.5]{ 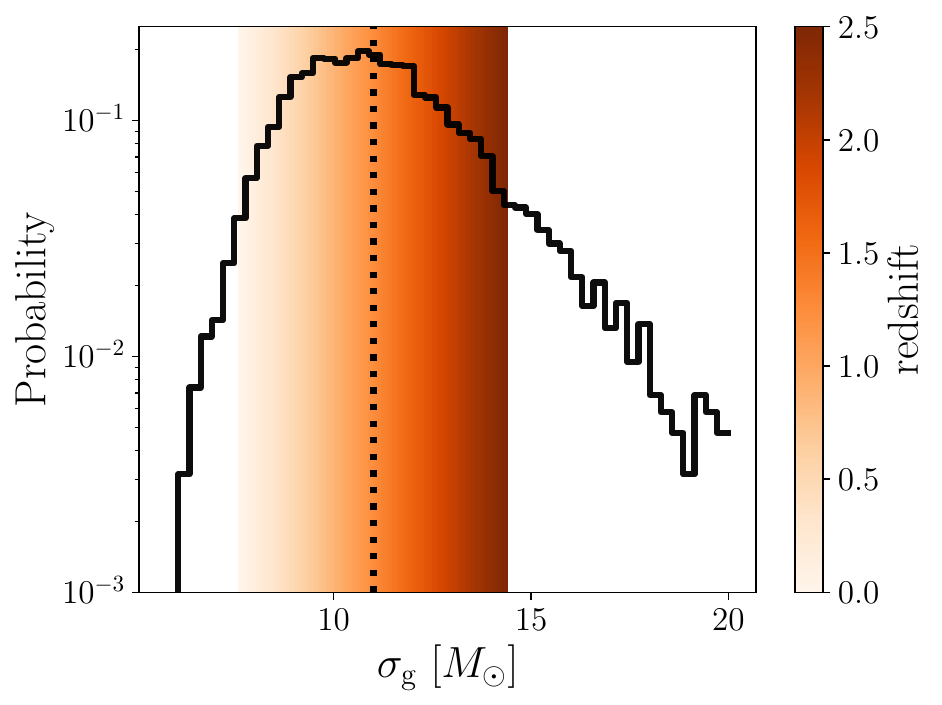}
    \caption{Posterior samples for $\sigma_g$ for the Redshift-Dependent scenario are plotted in \textit{black}. The dotted vertical line represents the median of the distribution. The colored region represents the values of the injected $\sigma_g$, $7.63+2.70z$, at different redshifts specified in the color bar.}
    \label{fig:sigma_g}
\end{figure}

\begin{figure}[]
    \centering
    \includegraphics[scale = 0.55]{ 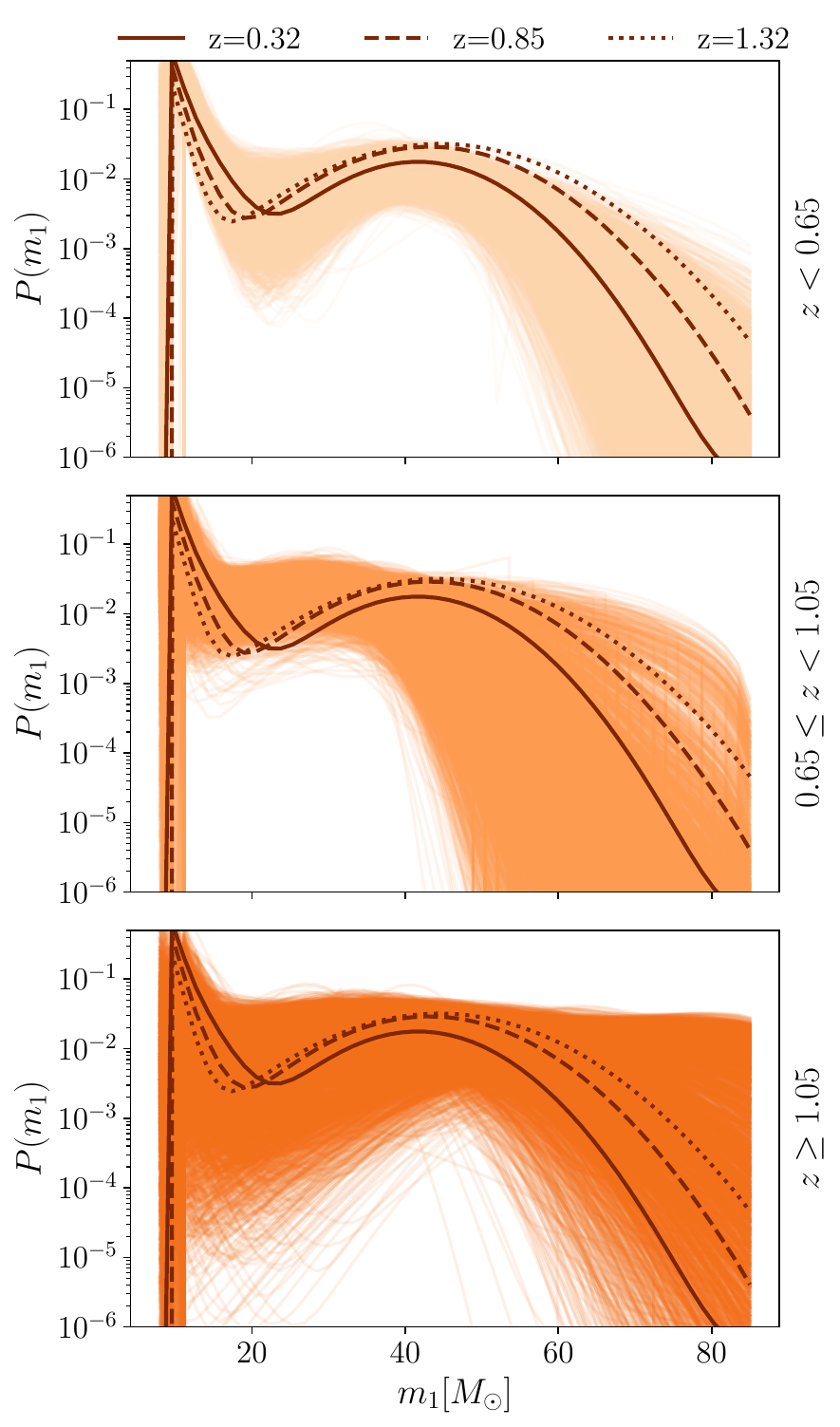}
    \caption{Posterior samples plots for the Redshift-Dependent scenario for the primary mass distribution following Eq.~\ref{eq:mass1_redshift_case}. In each plot, The colored lines show the posterior samples from the $68\%$ highest density interval. These samples have been categorized based on the redshift range of the events used for the analysis. The redshift range is shown on the left y-axis: the top panel represents events with $z<0.65$, the central panel shows events with $0.65\le z<1.05$, and the bottom panel corresponds to events with $z>1.05$. In each subplot, there are three darker lines (solid, dashed, and dotted) that remain consistent and represent the model at different redshift values, as defined in Tab.~\ref{tab:injections}. These redshift values have been selected to fall in the middle of the redshift ranges we divided the events into. The highest-redshift event is at $z=1.7$.}
    \label{fig:redshift_mass_samples}
\end{figure}

We perform a joint parameter estimation for the astrophysical population and cosmological parameters using \texttt{icarogw} on $80$ detected events in the mock data. Our main result for the Redshift-Dependent scenario is the one obtained with all 80 events and the full set of $10^6$ injections. The corresponding full corner plot with the parameters posteriors is shown in the appendix Fig.~\ref{fig:redshift_corner_full}. The main result we find is that the injected value of Hubble constant  {is outside $\sim 90\%$ C. I. of the inferred posterior distribution. A few other parameters related to the mass distribution of the GW sources} such as $\alpha, \beta, m_{\rm min}, \mu_{g}, \sigma_g, \lambda_{\rm peak}$ shows about 1-$\sigma$ discrepancy from the injected value. 

To explore the impact of redshift evolution of the mass distribution on the value of $H_0$, we show the joint contour of $H_0, \mu_g,$ and $\sigma_g$ in Fig. \ref{fig:redshift_red_corner}, and the comparison between the injected value of $\mu_{g}$ and $\sigma_g$ with the recovered posterior from all the events in Figs. \ref{fig:mu_g} and \ref{fig:sigma_g} respectively.  {These plots indicate that posterior on the parameters $\mu_{g, z}$ and $\sigma_z$ agrees with the injected values at different redshifts (shown by color-bar) within about $68\%$ C. I. of the inferred posterior distribution}. Furthermore, the samples within $68\%$ on the mass distribution for different redshift bins of the injected value are shown in Fig. \ref{fig:redshift_mass_samples} along with the injected distribution. We divided the population into three sets, redshift-wise (we chose $z<0.65$, $0.65\le z<1.05$, and $z\ge 1.05$ to ensure a similar number of events in each bin). We then compare the reconstructed posteriors with three reference models, evaluated at parameter values in the center of each redshift bin following Tab.~\ref{tab:injections}. In all cases, we used posterior samples within the $68\%$ highest density interval. These show a good agreement with the injected values in the simulation. 

\begin{figure}[]
    \centering
    \includegraphics[scale = 0.55]{ 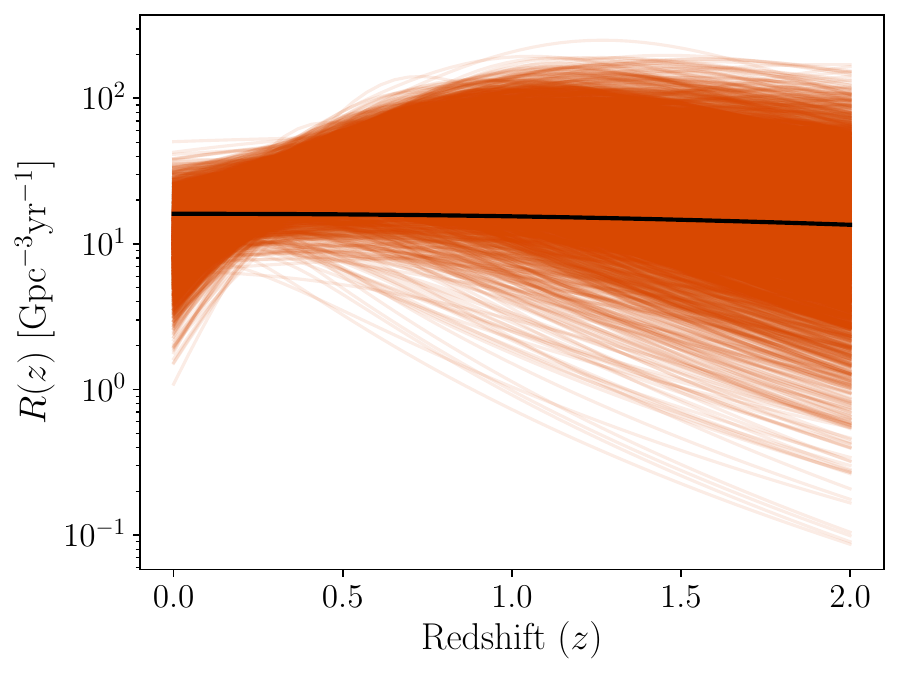}
    \caption{Same as in Fig.~\ref{fig:vanilla_rate_samples} but for the Redshift-Dependent scenario.}
    \label{fig:redshift_rate_samples}
\end{figure}

Among the parameters that control the merger rate such as $R_0, \gamma$, and $\kappa$,  the recovered values match well the injected value, as the underlying model is the same and there is no negative impact on the inference of the Hubble constant. The parameter $z_p$ which denotes the peak of the BBH merger rate distribution does not show a good recovery due to the fewer number of detected events at the injected value of high redshift ($z_p= 3.86$), as shown in the full corner plot in Fig.~\ref{fig:redshift_corner_full} (given in the appendix). The merger rate realizations from the samples of the posteriors with the $68\%$ C.I. are shown in Fig. \ref{fig:redshift_rate_samples}.

\begin{figure}[]
    \centering
    \includegraphics[scale = 0.5]{ 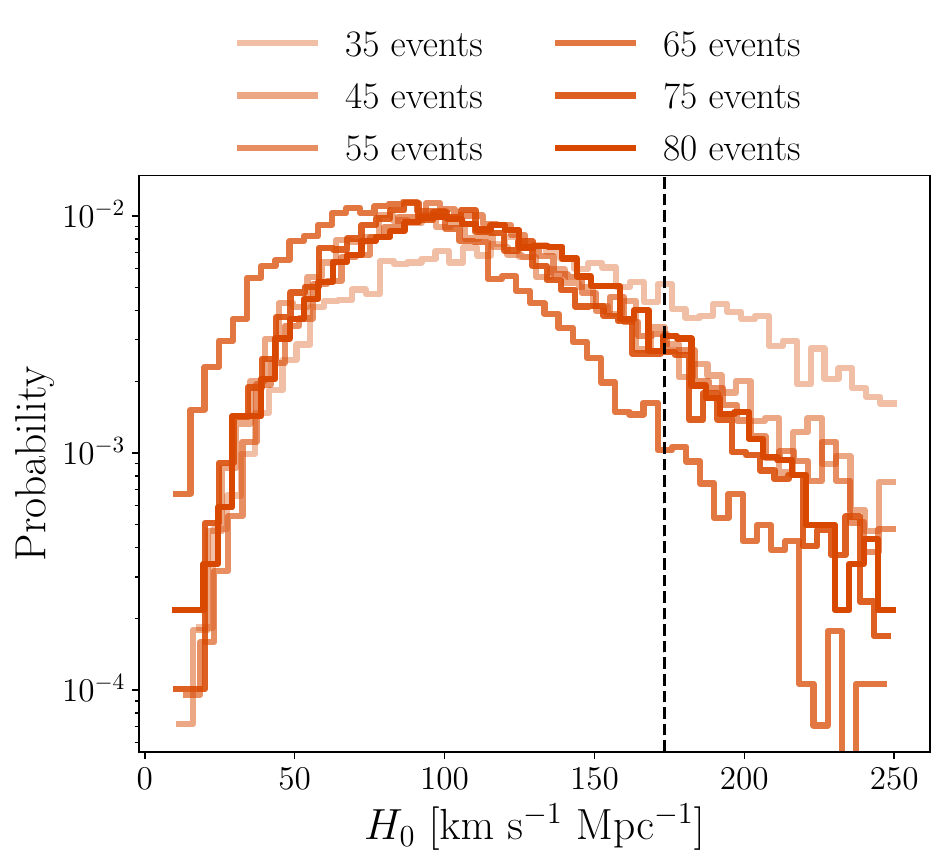}
    \caption{Check with the Redshift-Dependent scenario events. Posterior samples for the $H_0$ parameter using a different number of events. The dashed black line is the injected value.}
    \label{fig:red_ev_num}
\end{figure}

 {In Fig.~\ref{fig:red_ev_num}, we show the impact on $H_0$ from varying the number of events included in the analysis, specifically $[35, 45, 55, 65, 75, 80]$.} The distribution shows nearly consistent behavior between all the sub-samples  {having different number of events}. This indicates that the statistical uncertainty due to a fewer number of events is not causing any pronounced difference in the observed discrepancy in the $H_0$ posterior from the injected value shown by the black dashed line. For all the cases, the peak of the inferred $H_0$ posterior is  {at least} $76$ percentile away (towards a lower side) from the injected value. 

{We also performed statistical tests on the impact of the injection samples and dependence on event sample realization. For the former, we used $10^5$ and $10^6$ injection sets samples, and for the latter, we split the GW events into batches of 25 to analyze the resulting sub-populations. (For all statistical tests the results obtained are similar to the cases obtained for the vanilla case\footnote{The plots for these tests are shown in the \href{https://git.ligo.org/konstantin.leyde/cosmology_mdc_analysis_group_1_summary/-/tree/main/}{repository}.}.)}  {The impact of prior on the astrophysical parameters in inferring $H_0$ is also shown in the repository. The conclusion is similar to the Vanilla case, that the change in the prior mass distribution has a noticeable impact on $H_0$ posterior.}

\subsubsection{Discussion}\label{sec-dis}
In this section, we discuss our findings from the \texttt{Blinded-MDC} for the Redshift-Dependent scenario. As stated at the beginning of the motivation section, the reason for exploring a redshift-dependent mass model is to understand its impact on the inference of cosmology, due to ignorance of such evolution in reality due to complicated BBH formation channels and its dependence on parent star metallicity. As a result, we consider a mock simulation with redshift evolution in the mass and merger rate, where the merger rate evolution model agrees with the analysis model, but the mass distribution in the mock data differs from the one used in the analysis. The corresponding results obtained from the \texttt{Blinded-MDC} are presented in the previous section. 

The key result we are interested in exploring is to understand what is the impact of the mis-modeling on the inference of the Hubble constant. We find that the value of the Hubble constant is  {outside the $90\%$ C. I. from the injected value. The difference between the peak of the posterior distribution from the injected value is primarily driven by the correlation between the Hubble constant with the parameters $\mu_g$ and $\sigma_g$. The higher value of the inferred $\mu_g$ and $\sigma_g$ is driving a lower value of the Hubble constant. On comparing Fig. \ref{fig:vanilla_red_corner} and Fig. \ref{fig:redshift_red_corner} we find that the incorrect population assumption can drive the inference of the $\mu_g$ and $\sigma_g$ towards a higher value, resulting into a shift in the Hubble constant posterior from the injected value.} 

\begin{figure}[]
    \centering
    \includegraphics[scale = 0.45]{ 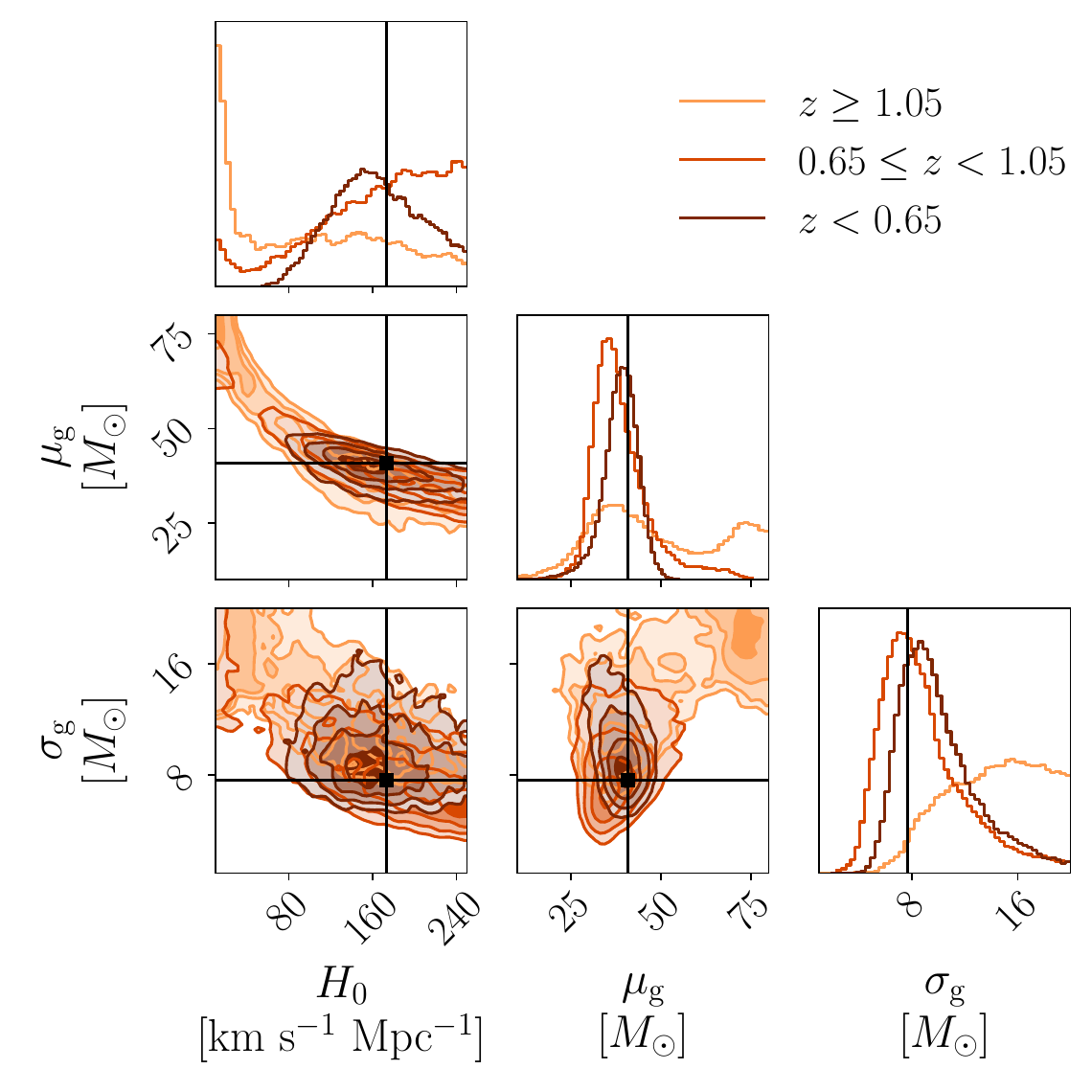}
    \caption{Reduced corner plot for $H_0$, $\mu_g$ and $\sigma_g$ parameters for the Redshift-Dependent scenario. We show results obtained selecting events in three different redshift ranges: at low redshift $z<0.65$, medium $0.65\le z<1.05$, and at redshifts higher than $z\ge 1.05$. The maximum redshift does not exceed $z=1.7$. The injected values are shown in \textit{black}. For $\mu_g$ and $\sigma_g$ they refer to the injected value at $z=0$ as in Tab.~\ref{tab:injections}. The three different colored regions of the 2D posterior represent, from the darkest to the lighter shade, respectively, the $39.3\%$, $86.5\%$, and $98.9\%$\,C.L. contours for a 2D Gaussian distribution. The dashed lines in the corresponding 1D histograms instead show the $68\%$\,C.L. of the marginalized distribution.}
    \label{fig:red_corner_zbins}
\end{figure}

To understand the origin of this discrepancy better, we scrutinize the results with events in three groups, divided based on their source redshift ({which are known from the simulation set}) as group-1 ($z \in \{ 0, 0.65\}$), group-2 ($z \in \{ 0.65, 1.05\}$), and group-3 ($z >1.05$) with a nearly equal number of sources in each group.  {The analysis in groups is performed by implementing the product over samples in Eq.~\eqref{eq:combined_posterior} in these three groups. The selection function of the analysis does not change for this sub-divided run, except for a reduction in T$_{\rm obs}$ by one-third for each group. This can be seen from Eq.~\eqref{eq:combined_posterior} that the term in the denominator, which captures the selection function, does not depend on the index (denoting individual events) over which the product in the numerator is carried out.}
The corresponding results on $H_0$ (after marginalizing over other parameters) are shown in Fig. \ref{fig:red_corner_zbins}. For the results from group 1, the value of $H_0$ matches very well with the injected value of the Hubble constant as shown by the solid black line. This is also valid for the other parameters $\mu_g$ and $\sigma_g$ which are strongly correlated with $H_0$. However, as we move towards higher redshift bins, i.e. group-2 and group-3, the posterior on $H_0$ moves towards a lower value, and the posteriors on $\mu_g$ and $\sigma_g$ moves
towards a higher value.  {The inference of the value of Hubble constant consistent with the injected value from the low redshift bin confirms that when there is only little redshift evolution of the mass distribution, the inference of the Hubble constant is correct and there is no confirmation bias. However, when the mass distribution is redshift evolving, the posterior distribution of the Hubble constant drives away from the injected value.}
 
This is arising because the sources contributing from the lowest redshift bin (in group-1) have the least impact on the redshift evolution of the mass distribution for the injection model discussed in Sec. \ref{sec:redshift-dependent}. As a result, when one is making an inference using a model with no redshift evolution, the error due to mis-modeling is minimal. On the other hand, for sources in group-2 and group-3, the redshift evolution of the $\mu_g$ and $\sigma_g$ parameter moves the values by about $10\%$ and $50\%$ respectively towards a higher value in comparison to the lowest redshift bin case, resulting into larger mis-modeling. As a result, the inferred value of the Hubble constant $H_0$ is shifted towards a lower value from the injected one, as shown in Fig. \ref{fig:red_corner_zbins}. The shift towards a lower value arises due to a strong negative correlation with the mass parameters ($\mu_g, \, \sigma_g$) and $H_0$. As the values at a higher redshift have intrinsically higher mass distribution, an analysis model that does not capture this effect associates this to a higher value of $\mu_g$ and $\sigma_g$, resulting in a lower value of inferred redshift than the true one\footnote{As the observed masses $m_z$ are redshifted, $m_z= m(1+z)$. An incorrect inference of true source frame mass will lead to an incorrect inference of redshift $z$ by $m_z/m$.}. As a result, the corresponding inferred $H_0$ is lower than the true value of the $H_0$ used in the simulation. This effect gets pronounced from group-2 to group-3 samples, as the change due to redshift in $\mu_g$ and $\sigma_g$ are large. 

 {To exclude that the lower $\rho_{\rm th}$ of the Vanilla case gives better constraints on the recovered hyperparameters due to a larger number of events, we ran a Vanilla analysis imposing a $\rho_{\rm th}=12$ as in the Redshift-Dependent scenario. The results, shown compared to the $\rho_{\rm th}=10$ ones in Fig.~\ref{fig:vanilla_full_corner_cfr_snr}, confirm the ones with the lower threshold, showing slightly enlarged errorbars, as a consequence of a reduced number of events ($164$ when $\rho_{\rm th}=12$). This check sets the Vanilla and the Redshift-Dependent analysis on the same footing, and corroborates our findings on the presence of a bias in the Redshift-Dependent scenario.}

\begin{figure*}[]
    \centering
    \includegraphics[scale = 0.525]{ 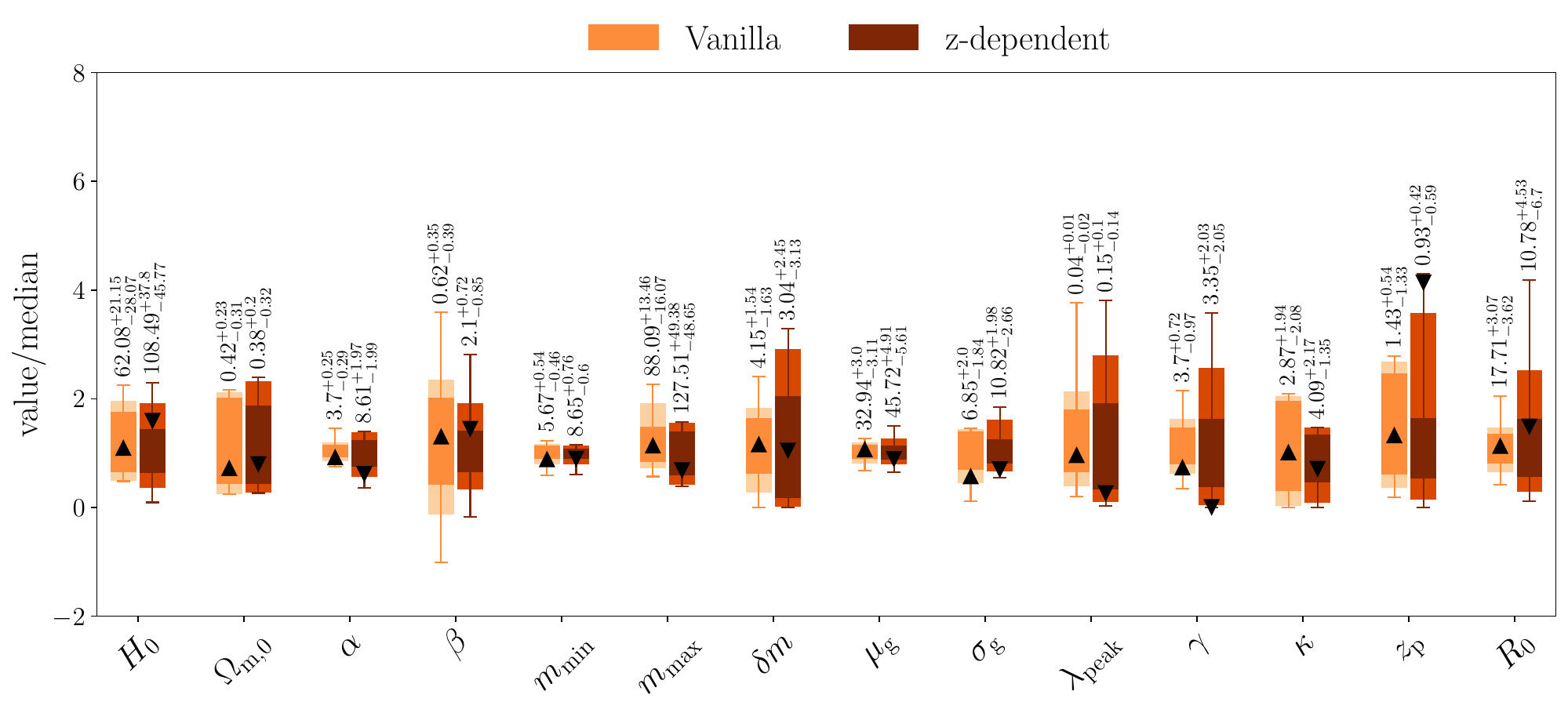}
    \caption{Summary plot of results for both the Vanilla (\textit{lighter shade}) and the Redshift-Dependent (\textit{darker shade}) models. The runs used all the available events (286 for the Vanilla case and 80 for the Redshift-Dependent one) with $10^6$ injections for selection effects. The $x-$axis represents the inferred hyperparameters, while the $y-$axis shows all the posterior samples divided by the median value of each distribution for easier comparison on a single plot. The boxplots are designed to encompass $1\sigma$ of the data around the median in the darker shade and up to $2\sigma$ with the lighter shade. The whiskers extend from the minimum to the maximum value to represent all the data. The black triangles represent the injected values divided by the corresponding median of the distribution. The text represents the median value of the parameter and the $68\%$ confidence interval around it. For the Redshift-Dependent model, the values shown in Tab.~\ref{tab:injections} (considering $z=0$) are being referred to.}
    \label{fig:summary}
\end{figure*}

The comparison of the marginalized posterior with respect to the injected value does not show any strong deviation for $\mu_g$ and $\sigma_g$ as shown in Fig. \ref{fig:summary} due to large uncertainties.   However, the value of $H_0$ starts showing deviation at $91$ percentile from the injected value due to the reasons mentioned above. This implies that even though the lowest distance bin value of $H_0$ does not show any discrepancy, a mild $10\%$ variation in $\mu_g$ can lead to noticeable deviation from the injected value due to the sources at higher distance. Though due to the blinded nature of the \texttt{Blinded-MDC}, in this analysis the value of the injected Hubble constant $H_0$ is large, the key finding showing the impact of redshift evolution of mass distribution will cause a discrepancy in the inferred value of $H_0$ remains valid. 

 {Along with the impact on the Hubble constant $H_0$, we also find that the inference of the BBH merger rate parameters such as $\gamma$ and $z_p$ also get impacted with a deviation from the injected value at about $2$-$\sigma$ or more, when the underlying population model of the mass distribution is the Redshift-Dependent scenario, and it is not taken into account in the analysis. Among these two parameter, the $z_p$ parameters show more deviation, as can be seen in Fig. \ref{fig:redshift_corner_full}. The constraints on the parameter $z_p$ is usually weak and shows a long tail due to the impact of selection function, which limits the detection of large number of sources from high redshifts. The parameters controlling the mass distribution are inferred within $1.5$-$\sigma$ from the injected value for both the Vanilla case and Redshift-Dependent scenario, with a maximum deviation noticed for the parameters $\alpha$ and $\sigma_g$, as shown in Fig. \ref{fig:redshift_corner_full}.}

A redshift evolution of the black hole mass distribution and also the merger rate is an expected outcome from an astrophysical perspective due to the dependencies of these on progenitor metallicity and SFR. The metallicity of the Universe varies by a few orders of magnitude with redshift, with a decreasing trend at high redshift. Furthermore, from galaxy to galaxy, there is a large variation in the metallicity. As a result, along with a monotonic redshift evolution, there is going to be additional stochastic behavior in the masses of BHs at a redshift. This can lead to a dispersive mass distribution from any monotonic behavior in redshift. In this \texttt{Blinded-MDC}, we have considered a simplistic scenario of linear evolution with redshift of the parameters controlling the mass distribution (as discussed in Sec. \ref{sec:redshift-dependent}). However, in reality, the mass distribution evolution can be even more severe. Future analysis will be conducted to understand this impact for a larger number of GW samples and better sensitivity for several astrophysical scenarios. 

 {The matter energy density $\Omega_{\rm m,0}$ is the other cosmological parameter inferred in our analyses. However, in all of our runs (see Fig.~\ref{fig:vanilla_corner_full} and Fig.~\ref{fig:redshift_corner_full}), show practically no constraining power on $\Omega_{\rm m,0}$.}

\section{Conclusion}\label{sec-conc}
This work presents the first \texttt{Blinded-MDC} for the inference of the cosmological parameters from GW observations feasible using BBHs from the currently ongoing network of LVK detectors with O4 sensitivity \citep{LIGOScientific:2014pky, VIRGO:2014yos, KAGRA:2013pob}. This study aims to explore the reliability of standard siren cosmology in inferring the true values of the cosmological parameters when the dark siren technique using a BBH mass distribution is used assuming an underlying model. Through this \texttt{Blinded-MDC} we explore both statistical and systematic uncertainty due to the interplay between astrophysical model assumptions on the BBH mass distribution and the inference of Hubble constant $H_0$ from the GW data. 
\begin{widetext}
    \begin{center}
    \begin{table}[h]
    \caption{Summary of the systematic and statistical tests performed in the analysis. }
        \label{tab:summary}
\begin{tabular}{ |c|c|c|c| } 
\hline
Type of Tests & Objective & Vanilla  {Scenario} & Redshift Dependent  {Scenario} \\
\hline
\multirow{4}{6em}{Systematic Uncertainties} & Inference of $H_0$ & not impacted &  impacted \\ 
& Impact of population prior on $H_0$ &  impacted  & impacted \\ 
& Inference of BBH mass distribution &  {mildly impacted} &  {mildly impacted} \\ 
& Inference of BBH Merger rate &  not impacted  &   {impacted} \\ 
\hline
\multirow{2}{6em}{Statistical Uncertainties} & Impact of number of samples in $p_{\rm det}$ & impacted &  impacted  \\ 
& Impact of sub-sample of events & mildly impacted & mildly impacted \\ 
\hline
\end{tabular}
    \end{table}
\end{center}
\end{widetext}

To explore this aspect, we considered two astrophysical scenarios namely, (i) the Vanilla model, which represents the currently known model of BBH mass distribution, and (ii) a Redshift-Dependent scenario which considers a mass distribution model with redshift evolution, which differs from the fiducial assumption. For both cases, GW mock samples are generated from the code \texttt{GWSIM} \citep{Karathanasis:2022hrb} by two independent simulation teams which are disjoint from the analysis team members. The key findings from this analysis are listed below: 
\begin{enumerate}
    \item For a given astrophysical model consistent with the Vanilla scenario, the cosmology inference code can infer the injected value for all the parameters correctly in a blinded setup using the mass distribution (See Fig. \ref{fig:vanilla_red_corner}).
    \item The impact of fewer samples in the injection set for the calculation of the detection probability mildly overestimates (underestimates) the posterior on the Hubble Constant $H_0$ at the lower (higher) values in comparison to a case with more samples (See Fig. \ref{fig:vanilla_inj_set}). 
    \item If the mass distribution of the BBHs shows mild redshift evolution (as shown in Fig.~\ref{fig:Pz1}) which is theoretically unknown (or unknown from independent observations) due to the evolution of stellar properties with redshift,  {then the inferred value of the Hubble constant shows discrepancy from the injected value at more than $1\sigma$}  for O4 noise sensitivity (See Fig. \ref{fig:redshift_red_corner}). 
    \item For the case of redshift evolving mass distribution, the observed statistical fluctuation in the inferred value of the parameters (in particular the value of $H_0$) is not significant in comparison to the systematic error due to mass mis-modelling (See Fig. \ref{fig:red_ev_num}). 
    \item The inferred parameters describing the astrophysical mass distribution do not show any significant departure from the injected parameters of the mass distribution when the underlying true model is unknown. As a result, even if the mass distribution evolves mildly (considered in this analysis),  {the values of $\mu_g$ and $\sigma_g$ agree well with the inferred posterior distribution} (See Figs. \ref{fig:mu_g} and \ref{fig:sigma_g}). 
    
    \item The last point implies that in an analysis performed in different redshift bins of the GW sources, the mass distribution in individual redshift bins does not show any significant deviation from the injected value due to large error bars. However, the combined posterior on the Hubble constant leads to a noticeable discrepancy from the true value (See Fig. \ref{fig:red_corner_zbins}).  

    \item  {For the Redshift-dependent case, the inference of some of the parameters which controls the merger rate namely $\gamma$ and $z_p$ gets partially impacted when the underlying population model is unknown. We find that both these parameters are impacted by around $2-\sigma$ or more. However, it needs to be pointed out that value of $z_p$ is poorly constrained also due to selection effect arising from the detection of fewer sources at high redshift.}  
    \end{enumerate}

This study indicates that the analysis method and setup of the code work well when the model used in the mock data generation is the same as the one used in the analysis. But if there are departures in the mass model that are redshift dependent, then even though the impact on the inference of the astrophysical population model is not pronounced, it becomes more pronounced on the inference of $H_0$. This arises because the value of $H_0$ gets constrained by all the sources located at different redshift ranges. If there is a monotonic evolution (increase or decrease) in the mass distribution of the BBHs, then the inferred redshift starts to differ from the true value of the redshift in a monotonic way. As a result, the corresponding value of the Hubble constant $H_0$ will monotonically decrease or increase. In reality, inferred mass distribution with a wrong model may still give a reasonable astrophysical understanding of the mass distribution, but a value of $H_0$ towards a lower (or higher) value from the true value, depending on whether the mass distribution is increasing (or decreasing) with redshift. The findings from the \texttt{Blinded-MDC} obtained here have a crucial impact on the robustness of standard siren cosmology when the astrophysical assumptions on mass distribution are made. As expected, the discrepancy gets pronounced with a higher number of GW events. 

In the future, it will be crucial to understand the interplay between the astrophysical assumptions and cosmological parameters with better sensitivity and a higher number of events. It will be important to investigate the robustness in inferring cosmological parameters for different astrophysical scenarios of BBH formation, depending on their delay time distribution, and stellar metallicity. Along with these, understanding the robustness of other dark sirens techniques related to galaxy incompleteness and the assumptions on host galaxy luminosity for the statistical host techniques will be necessary \citep{Perna:2024lod, Hanselman:2024hqy}. Similarly, for the cross-correlation technique, understanding the impact on $H_0$ inference due to the redshift dependence of the GW bias parameter \citep{Diaz:2021pem, Dehghani:2024wsh}. In summary, this study performs the first \texttt{Blinded-MDC} and provides a framework to check the reliability of cosmological inferences from the standard sirens cosmology for the scenarios of dark sirens, which can be extended for other techniques as well. To make a robust measurement of $H_0$ from standard sirens which can shed light on the ongoing \textsc{$H_0$- Tension} \citep{Verde:2019ivm,Abdalla:2022yfr}, \texttt{Blinded-MDC} of the analysis techniques will be crucial to explore possible sources of modeled and un-modeled systematic uncertainties. We will explore different such scenarios in future  \texttt{Blinded-MDC} analyses.

\textbf{Acknowledgment}
The authors are very thankful to Archisman Ghosh for reviewing the manuscript as a part of the LIGO Publication and Presentation policy and providing valuable comments on the draft. The authors also thank to the LIGO-Virgo-KAGRA Scientific Collaboration for providing noise curves. The authors are grateful for computational resources provided by the LIGO Laboratory and supported by National Science Foundation Grants PHY-0757058 and PHY-0823459 and the computing resources of the \texttt{⟨data|theory⟩ Universe-Lab}, supported by TIFR and the Department of Atomic Energy, Government of India. LIGO, funded by the U.S. National Science Foundation (NSF), and Virgo, supported by the French CNRS, Italian INFN, and Dutch Nikhef, along with contributions from Polish and Hungarian institutes. This collaborative effort is backed by the NSF’s LIGO Laboratory, a major facility fully funded by the National Science Foundation. The research leverages data and software from the Gravitational Wave Open Science Center, a service provided by LIGO Laboratory, the LIGO Scientific Collaboration, Virgo Collaboration, and KAGRA. Advanced LIGO's construction and operation receive support from STFC of the UK, Max-Planck Society (MPS), and the State of Niedersachsen/Germany, with additional backing from the Australian Research Council. Virgo, affiliated with the European Gravitational Observatory (EGO), secures funding through contributions from various European institutions. Meanwhile, KAGRA's construction and operation are funded by MEXT, JSPS, NRF, MSIT, AS, and MoST. This material is based upon work supported by NSF’s LIGO Laboratory which is a major facility fully funded by the National Science Foundation. This work is supported in part by the Perimeter Institute for Theoretical Physics. Research at Perimeter Institute is supported by the Government of Canada through the Department of Innovation, Science and Economic Development Canada and by the Province of Ontario through the Ministry of Economic Development, Job Creation and Trade. MCE is supported by COSMOGRAV. The research of FB and CT is supported by Ghent University Special Research Funds (BOF) project BOF/STA/202009/040, the inter-university iBOF project BOF20/IBF/124, and the Fonds Wetenschappelijk Onderzoek (FWO) research project G0A5E24N. They also acknowledge support from the FWO International Research Infrastructure (IRI) grant I002123N for Virgo collaboration membership and travel to collaboration meetings. The work of SM and MRS is part of the \texttt{⟨data|theory⟩ Universe-Lab}, supported by TIFR and the Department of Atomic Energy, Government of India.  AER was supported by the UDEA projects
 2021-44670, 2019-28270, 2023-63330. RG is suported by STFC grant ST/V005634/1. This project has received financial support from the CNRS through the AMORCE funding framework and from the Agence Nationale de la Recherche (ANR) through the MRSEI project ANR-24-MRS1-0009-01, Spanish Research Project PID2021-123012NB-C43 [MICINN-FEDER], and Centro de Excelencia Severo Ochoa Program CEX2020-001007-S at IFT.

\bibliography{references.bib}{}
\bibliographystyle{aasjournal}

\appendix
\section{\bilby{} runs for events PE}
We used 286 events and 80 binary black hole events for the Vanilla and the Redshift-Dependent scenarios, respectively. The population models used to generate the events are described in Sec.~\ref{sec-pop}. We employed \bilby{} \citep{Ashton_2019} to obtain the parameter estimation for all of these events. The two models' events analysis settings are common and are described in the following. 

As stated in the main text, the analysis was done for the $\mathcal{M}$, q, $d_L$, Dec, RA, $i$, $\Psi$, $\phi$, and $t_c$ parameters (see a brief description in Sec.~\ref{sec-pop}). The remaining spin parameters (6 in total) were fixed to zero with a $\delta-$prior. We used the \texttt{dynesty} sampler and the \texttt{IMRPhenomPv2} waveform approximant.  {This waveform does not include higher-order harmonics. While it does incorporate spin precession effects, we did not vary the spin parameters in our analysis. Although both higher-order modes and spin precession can help break degeneracies between parameters, their impact on detection and cosmological inference is expected to be limited given the current detector sensitivity. Therefore, we adopted this simplified waveform setup to reduce computational complexity}.The detectors settings are outlined in Tab.~\ref{tab:duty}. For the Vanilla scenario, we considered 3 years of observations corresponding to a combination of the first four observing runs. Instead, we limited our analysis to the fourth observing run corresponding to 1 year of observing time for the redshift-dependent case. 

\begin{figure*}
    \centering
    \includegraphics[scale = 0.35]{ 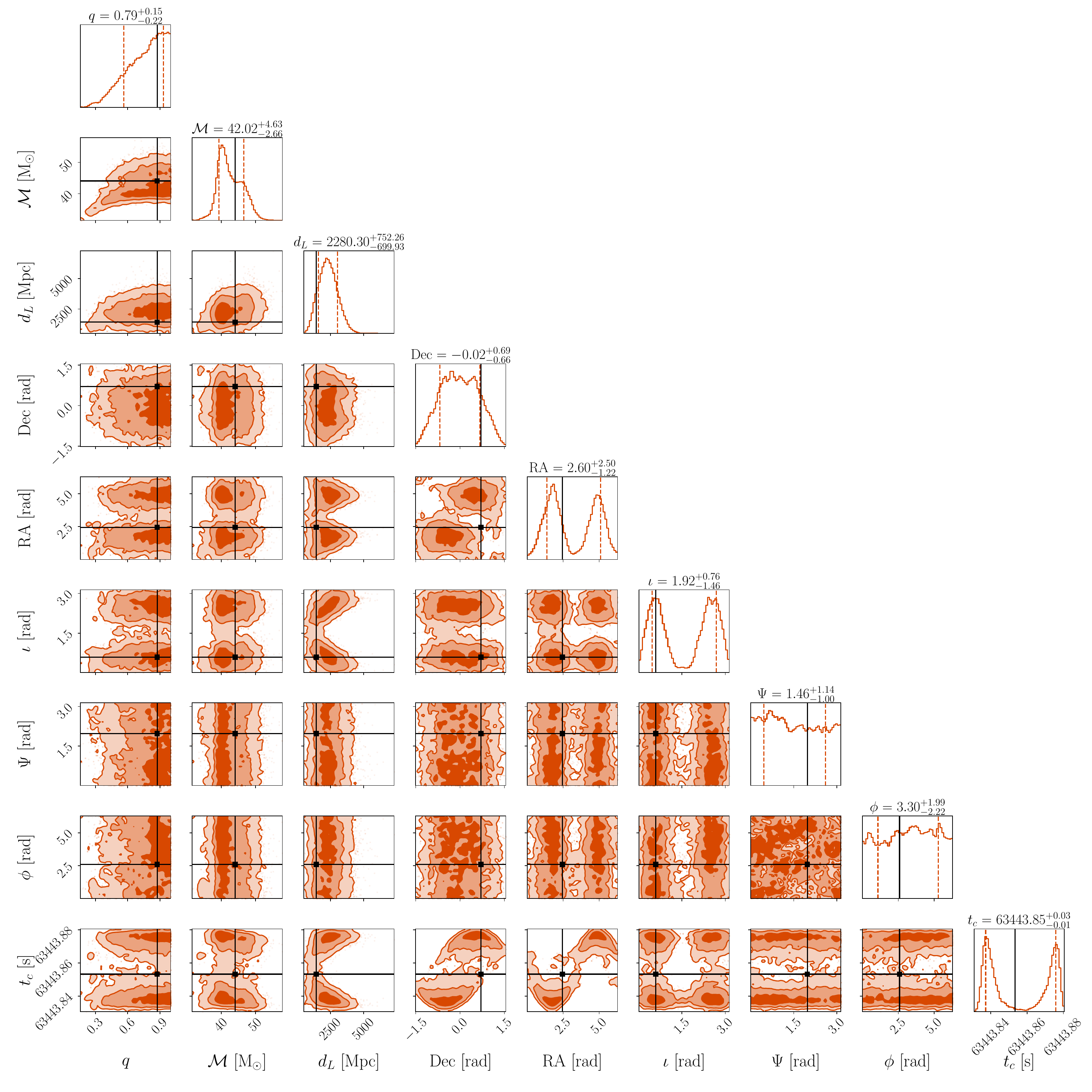}
    \caption{Posterior samples from the \bilby{} run for one event from the Vanilla case. The injected values are indicated in \textit{black}. The three different colored regions of the 2D posterior represent, from the darkest to the lighter shade, respectively, the $39.3\%$, $86.5\%$, and $98.9\%$\,C.L. contours for a 2D Gaussian distribution. The dashed lines in the corresponding 1D histograms instead show the $68\%$\,C.L. of the marginalized distribution.}
    \label{fig:bilby_vanilla}
\end{figure*}

\begin{figure*}
    \centering
    \includegraphics[scale = 0.35]{ 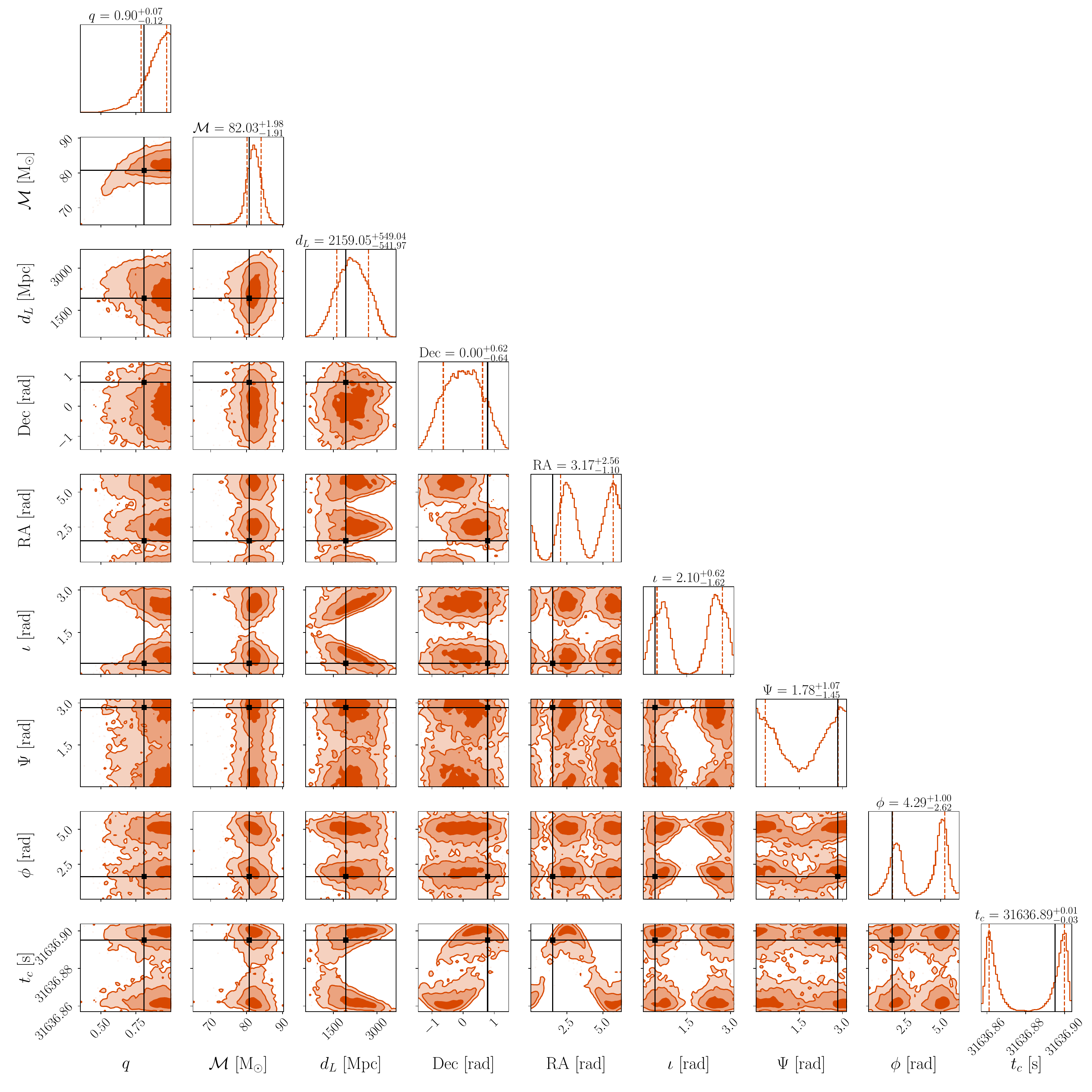}
    \caption{Posterior samples from the \bilby{} run for one event from the Redshift-Dependent scenario. The injected values are indicated in \textit{black}. The three different colored regions of the 2D posterior represent, from the darkest to the lighter shade, respectively, the $39.3\%$, $86.5\%$, and $98.9\%$\,C.L. contours for a 2D Gaussian distribution. The dashed lines in the corresponding 1D histograms instead show the $68\%$\,C.L. of the marginalized distribution.}
    \label{fig:bilby_redshit}
\end{figure*}

In Fig.~\ref{fig:bilby_vanilla} and Fig.~\ref{fig:bilby_redshit}, we show the parameter estimation results for the Vanilla and the Redshift-Dependent scenarios, respectively. The posterior samples for the 9 parameters considered in the analysis are shown in \textit{orange}, and the true (injected) values are indicated with a \textit{black} line. We chose one event per scenario as a reference example.

\section{Full hyper-parameters posteriors}
\label{app:full_posteriors}
\begin{figure*}[]
    \centering
    \includegraphics[scale = 0.245]{ 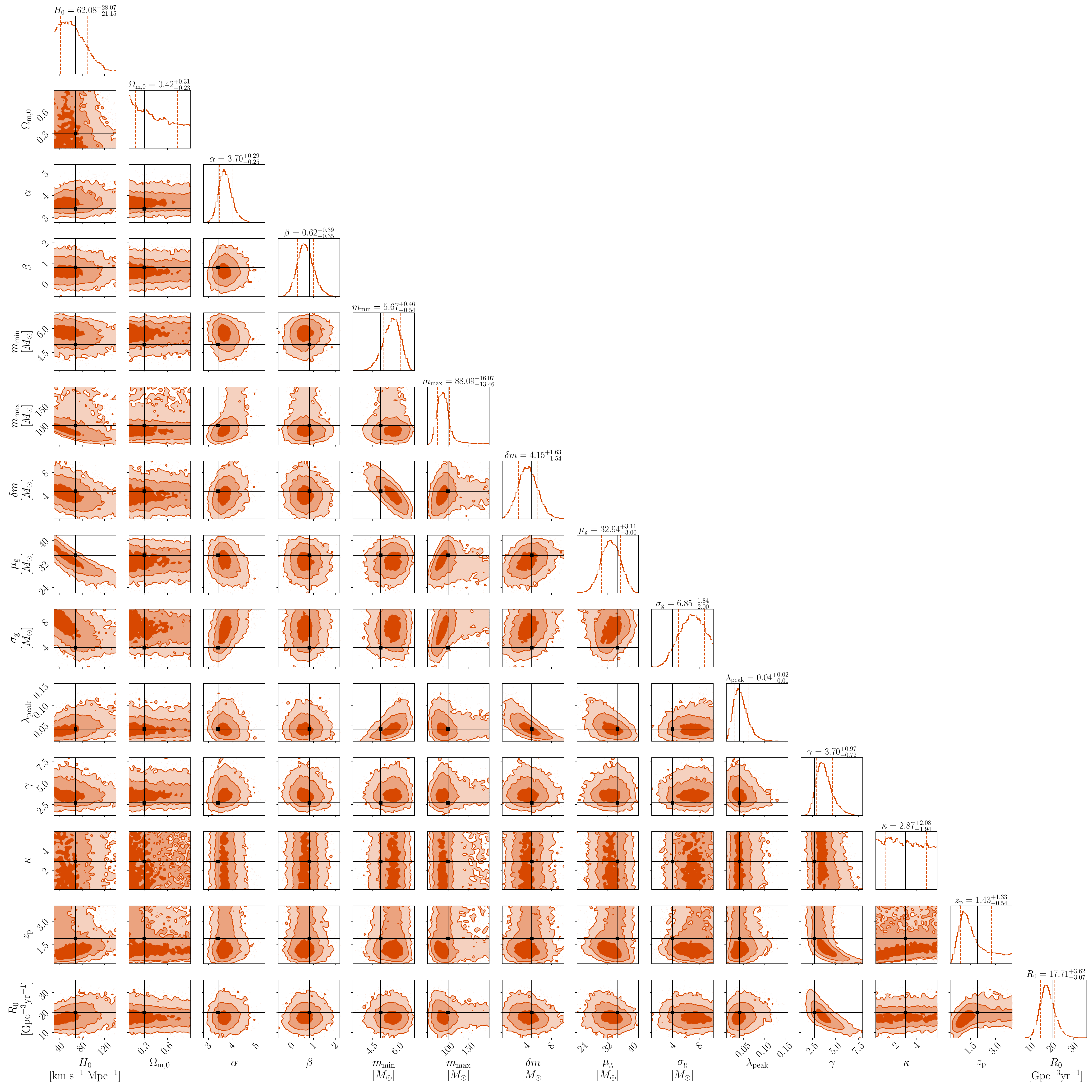}
    \caption{Posterior samples for the Vanilla scenario, obtained by analysing all the 286 events using $10^6$ injections to estimate the detection probability $p_{\rm det}$. The three different colored regions of the 2D posterior represent, from the darkest to the lighter shade, respectively, the $39.3\%$, $86.5\%$, and $98.9\%$\,C.L. contours for a 2D Gaussian distribution. The dashed lines in the corresponding 1D histograms instead show the $68\%$\,C.L. of the marginalized distribution.}
    \label{fig:vanilla_corner_full}
\end{figure*}

\begin{figure*}[]
    \centering
    \includegraphics[scale = 0.245]{ 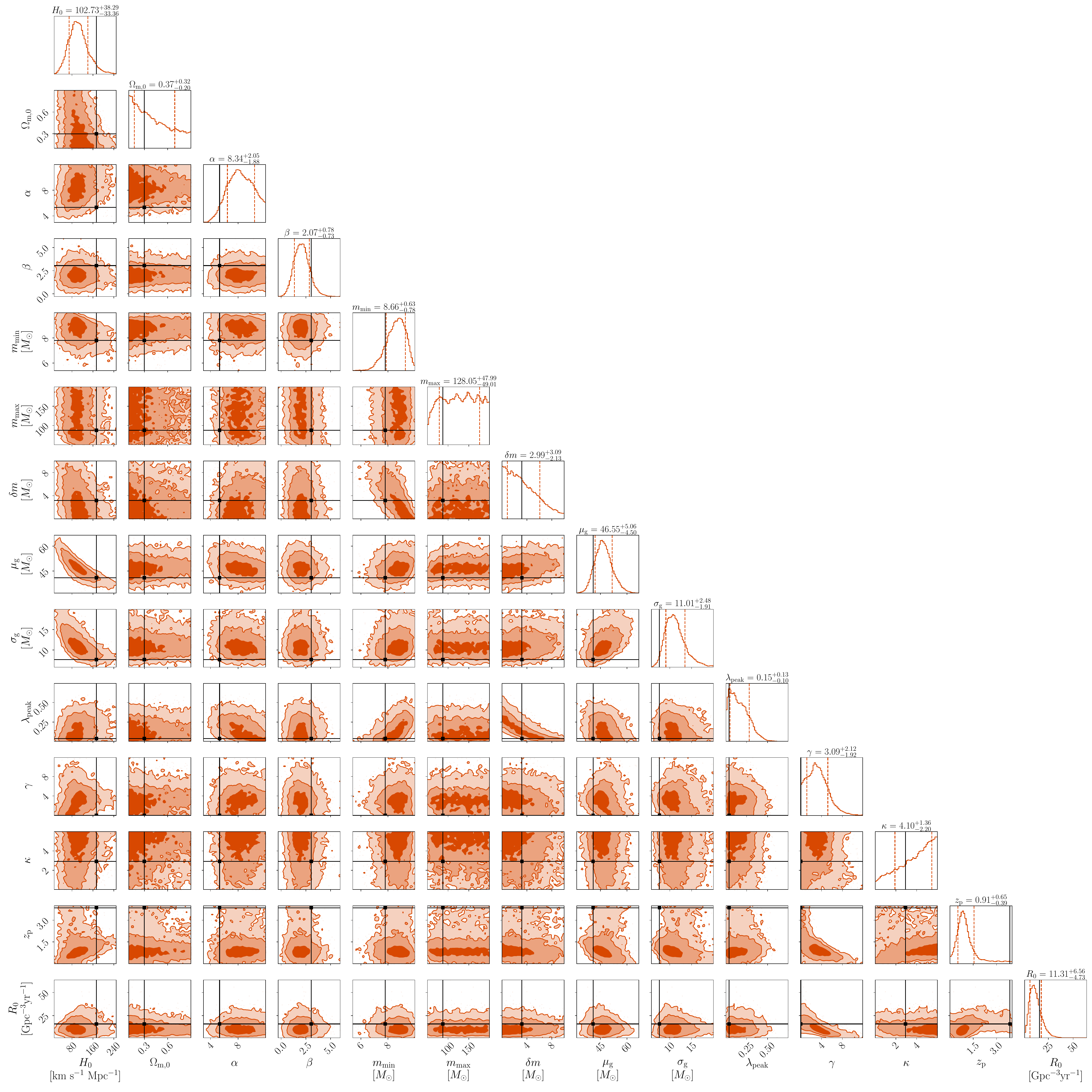}
    \caption{Posterior samples for the Redshift-Dependent scenario, obtained by analysing all the 80 events using $10^5$ injections to estimate the detection probability $p_{\rm det}$. The three different colored regions of the 2D posterior represent, from the darkest to the lighter shade, respectively, the $39.3\%$, $86.5\%$, and $98.9\%$\,C.L. contours for a 2D Gaussian distribution. The dashed lines in the corresponding 1D histograms instead show the $68\%$\,C.L. of the marginalized distribution.}
    \label{fig:redshift_corner_full}
\end{figure*}

In this appendix, we report the full corner plots of the hyper-parameters in both the Vanilla and the Redshift-Dependent scenarios, shown in Fig.~\ref{fig:vanilla_corner_full} and in Fig.~\ref{fig:redshift_corner_full}, respectively. For both scenarios, we used all the selected events( 286 for the Vanilla case and 80 events for the Redshift-Dependent), and use $10^6$ injections for the selection effects computation.

\begin{figure*}[]
    \centering
    \includegraphics[scale = 0.245]{ 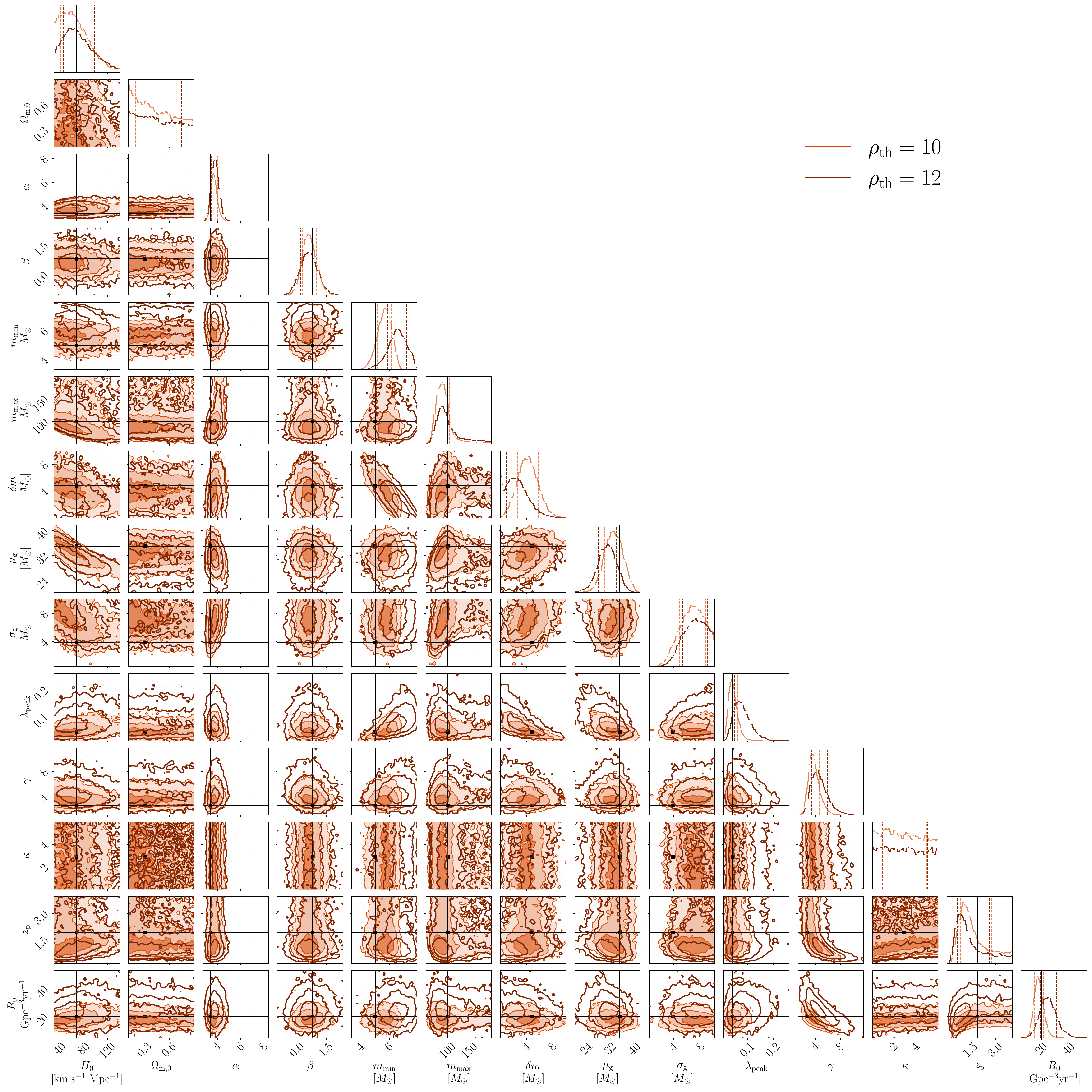}
    \caption{\bf Posterior samples for the Vanilla scenario, obtained by imposing a threshold $\rho_{\rm th}$ of 12 as in the Redshift-Dependent scenario, compared to the $\rho_{\rm th}=10$ as in Fig.~\ref{fig:vanilla_corner_full}. In the higher threshold case, we use $10^5$ injections to estimate the detection probability $p_{\rm det}$. The total amount of events amounts to 164 in this case. The three different colored regions (full for $\rho_{\rm th}=10$, and contours only for $\rho_{\rm th}=12$, of the 2D posterior represent, from the darkest to the lighter shade, respectively, the $39.3\%$, $86.5\%$, and $98.9\%$\,C.L. contours for a 2D Gaussian distribution. The dashed lines in the corresponding 1D histograms instead show the $68\%$\,C.L. of the marginalized distribution.}
    \label{fig:vanilla_full_corner_cfr_snr}
\end{figure*}

 {In Fig.~\ref{fig:vanilla_full_corner_cfr_snr}, we report the full corner plot with a $\rho_{\rm th}=12$, which aligns with the criteria used for the Redshift-Dependent scenario. On the same figure, we also report the $\rho_{\rm th}=10$ results for an easier comparison.}

\section{Additional analyses of the vanilla model}
\label{app: additional results vanilla model}
\begin{figure}[]
    \centering
    \includegraphics[scale = 0.5]{ 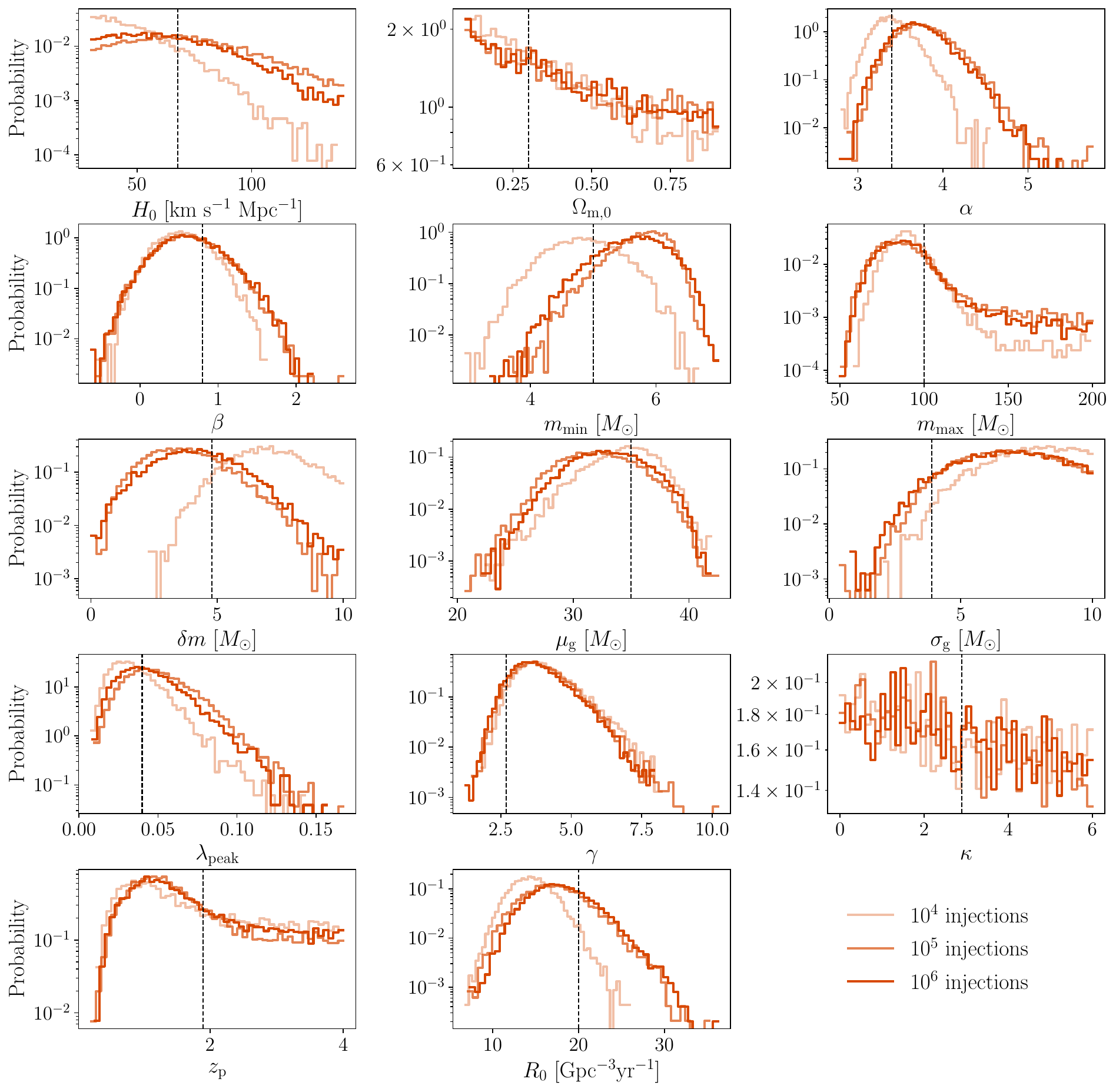}
    \caption{Vanilla case test on the dependence on the number of injections.}
    \label{fig:vanilla_full_posteriors_inj_set}
\end{figure}

\begin{figure}[]
    \centering
    \includegraphics[scale = 0.5]{ 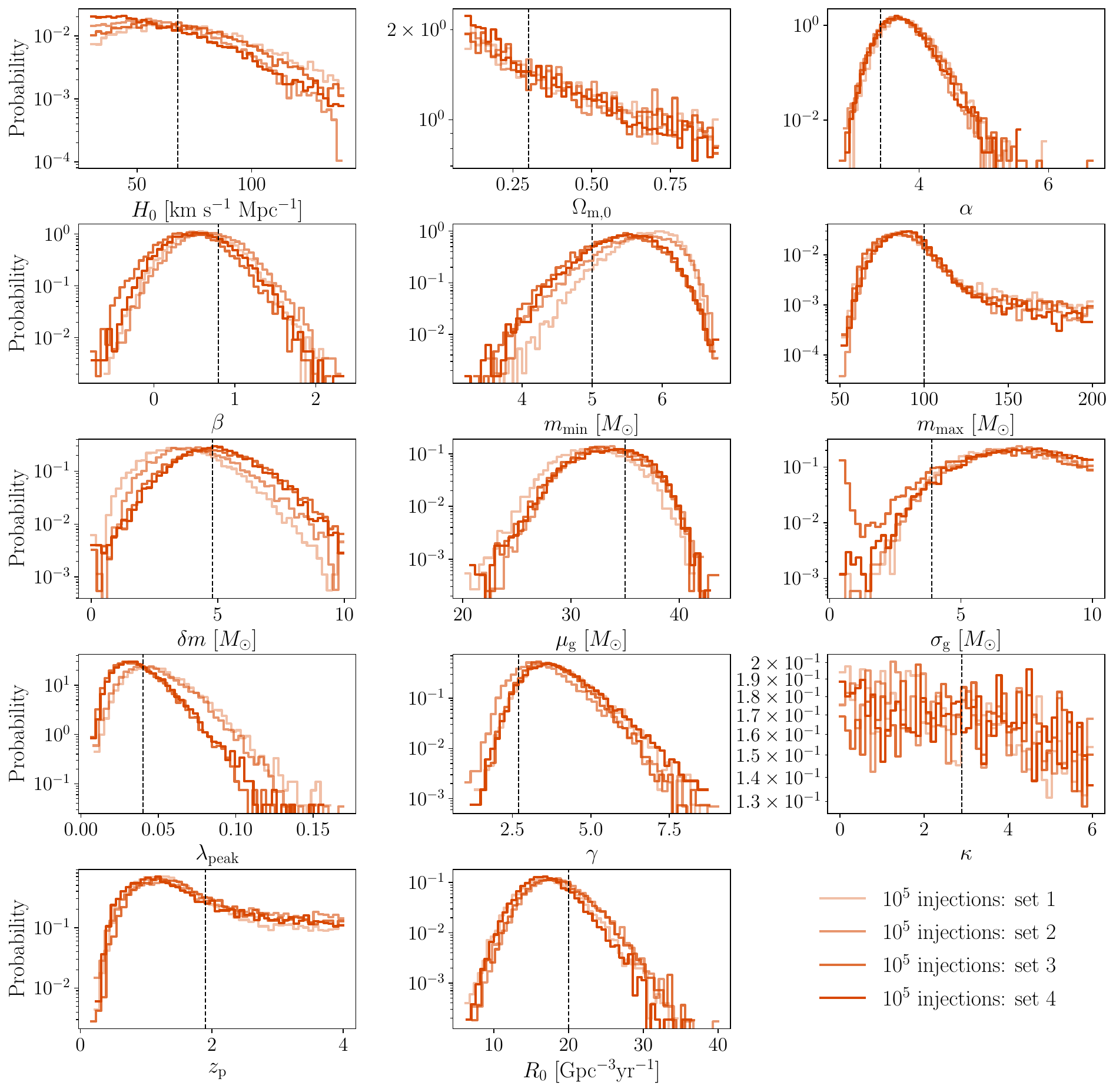}
    \caption{ {Vanilla case test on the dependence on the sample of injections, which number is fixed to $10^5$.}}
    \label{fig:vanilla_full_posteriors_inj_sample}
\end{figure}

\begin{figure}[]
    \centering
    \includegraphics[scale = 0.5]{ 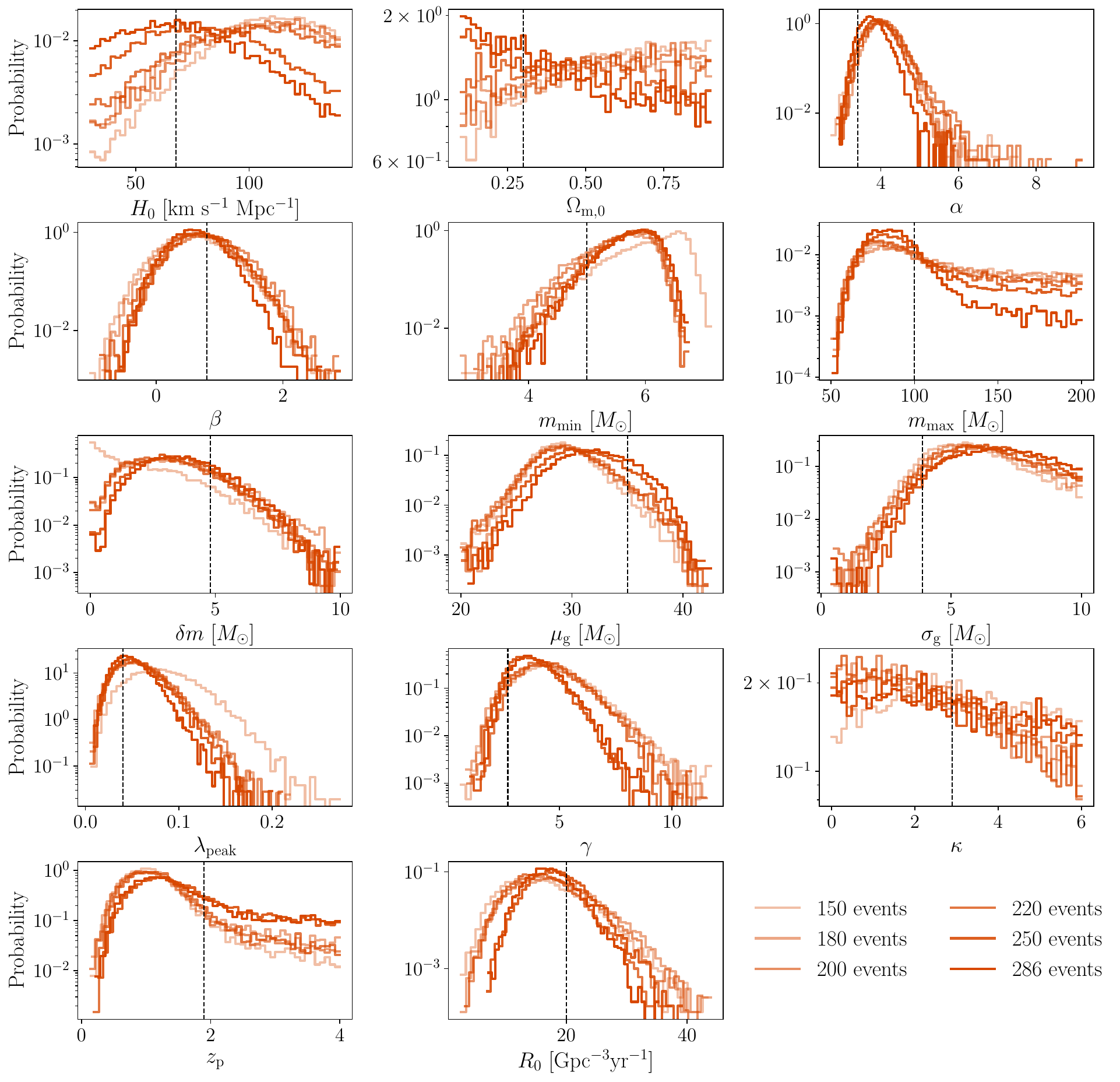}
    \caption{Vanilla case test on the dependence on the number of events.}
    \label{fig:vanilla_full_posteriors_ev_num}
\end{figure}

\begin{figure}[]
    \centering
    \includegraphics[scale = 0.5]{ 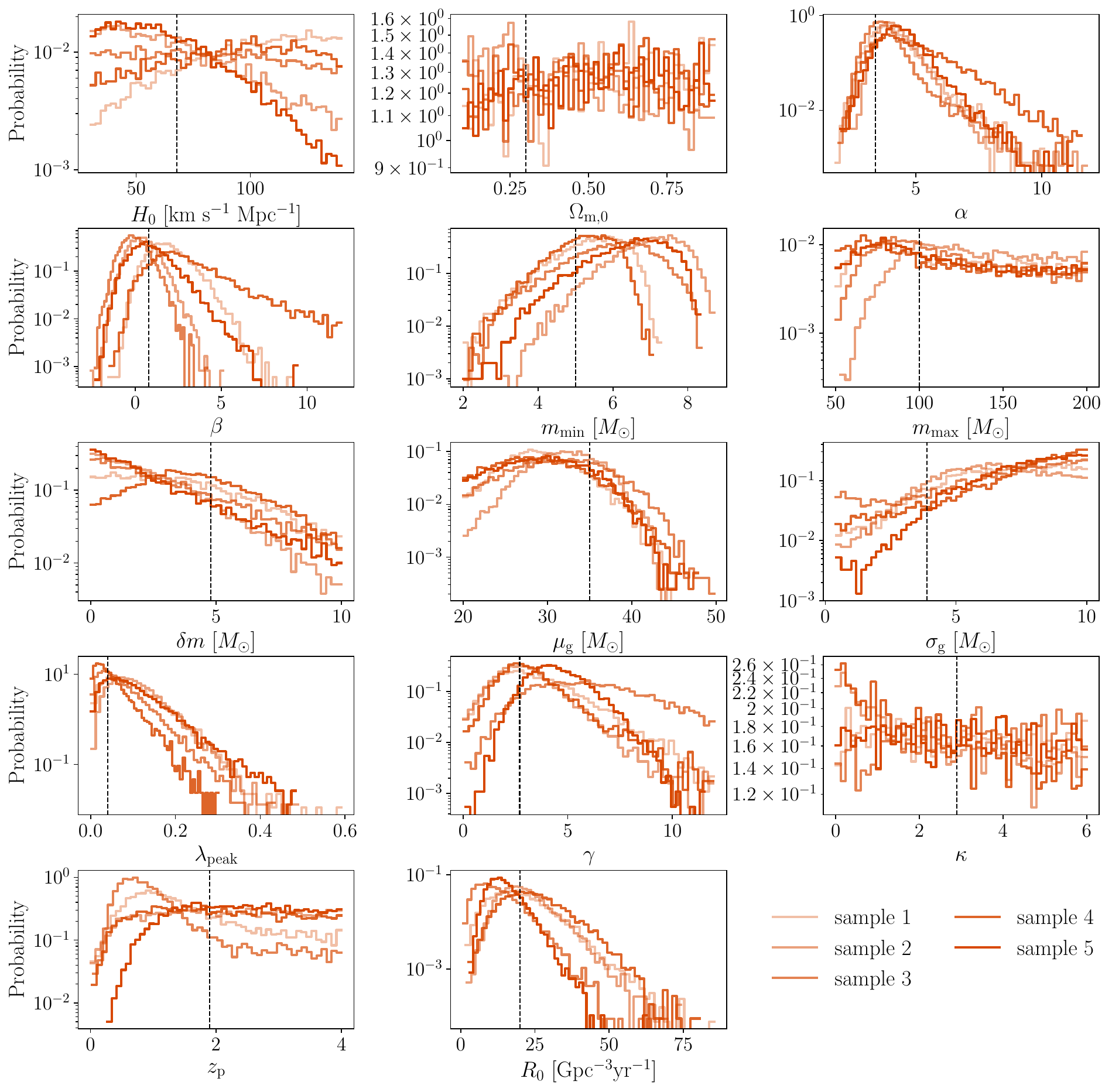}
    \caption{Vanilla case test on the dependence on the sample of events, keeping the number of events fixed to $50$ in each sample.}
    \label{fig:vanilla_full_posteriors_pe_sample}
\end{figure}

\begin{figure}[]
    \centering
    \includegraphics[scale = 0.5]{ 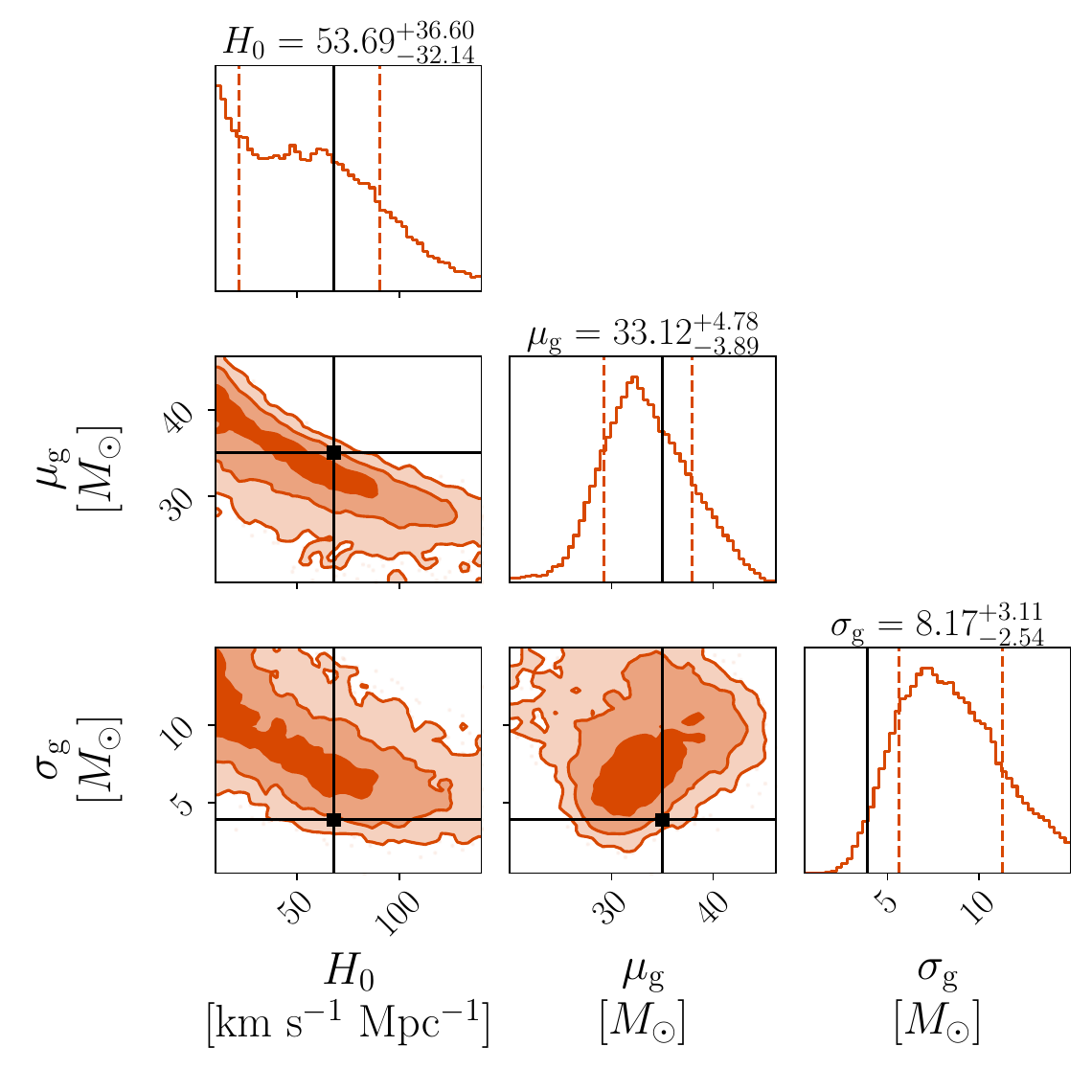}
    \caption{ {Vanilla case test with larger prior on $H_0,\, \mu_g$ and $\sigma_g$ parameters. The injected values are represented by the solid black lines. The three different colored regions of the 2D posterior represent, from the darkest to the lighter shade, respectively, the $39.3\%$, $86.5\%$, and $98.9\%$\,C.L. contours for a 2D Gaussian distribution. The dashed lines in the corresponding 1D histograms instead show the $68\%$\,C.L. of the marginalized distribution.}}
    \label{fig:red_vanilla_corner_larger_priors}
\end{figure}

We report the full plots for the statistical tests done for the Vanilla case. We refer to Sec.~\ref{sec-results} for the detailed description of the tests. In summary:
\begin{itemize}
    \item In Fig.~\ref{fig:vanilla_full_posteriors_inj_set}, we show that changing the number of injections to evaluate selection effects between $10^4,\,10^5$ and $10^6$ gives different results for all the hyper-parameters.
    \item  {In Fig.~\ref{fig:vanilla_full_posteriors_inj_sample}, we show that changing the realization of the injection sample while keeping the number of injections fixed at $10^5$ does not have any effect on the results.}
    \item In Fig.~\ref{fig:vanilla_full_posteriors_ev_num}, we show the impact of choosing a different number of events in the analysis. Apart from $H_0$ and $\Omega_{\rm m, 0}$, which are slightly affected, all the other hyper-parameters do not show major discrepancies if the number of events is larger than 180. We notice that with 150 events only we have some deviations from the general trend.
    \item In Fig.~\ref{fig:vanilla_full_posteriors_pe_sample}, we show that changing the set of events with a total number fixed to 50, does not result in major discrepacies. This confirms that we do not have any particular outlier in our dataset.
    \item  {In Fig.~\ref{fig:red_vanilla_corner_larger_priors} with show a companion plot to Fig.~\ref{fig:vanilla_red_corner}, where we did the analysis using a larger prior on both the $H_0$ and the $\sigma_g$ parameters to see if this would have reduced the railing of $H_0$ on the left and the one of $\sigma_g$ on the right. We see that we are getting less support for $\sigma_g$ at higher values, but still the strong correlation with $H_0$ allows for low values for the Hubble constant.}
\end{itemize}

\end{document}